\documentclass[a4paper,12pt]{article}
\usepackage[left=2cm,right=2cm,top=2cm,bottom=2cm]{geometry}
\usepackage[usenames,dvipsnames,svgnames,table]{xcolor}
\usepackage{psfrag,amsmath,amsfonts,amssymb,latexsym,amsthm,verbatim,color,lscape,multirow,graphicx,subcaption}
\usepackage[figuresright]{rotating}
\usepackage[sort&compress,numbers]{natbib}
\usepackage{url}
\usepackage{hyperref}
\hypersetup{colorlinks,
citecolor=black,
filecolor=black,
linkcolor=black,
urlcolor=black,
pdftex}
\usepackage{pdfpages}
\usepackage{booktabs}

\usepackage{colortbl}
\definecolor{gray}{rgb}{0.9,0.9,0.9}

\usepackage{times}
\usepackage{blindtext}
\usepackage[font={footnotesize,it}]{caption}

\usepackage{sectsty}  
\subsectionfont{\normalsize\bf}
\sectionfont{\large\bf}

\usepackage[symbol]{footmisc}
\renewcommand{\thefootnote}{\fnsymbol{footnote}}

\usepackage{algorithm}
\usepackage[noend]{algpseudocode}

\def\x{\mathbf{x}}

\def\s{\sigma}

\def\X{\mathbf{X}}

\def\a{\alpha}
\def\g{\gamma}
\def\b{\beta}
\def\de{\delta}

\def\varpibf{\boldsymbol{\varpi}}

\def\th{\theta}
\def\thbf{\boldsymbol{\theta}}

\def\Sbf{\boldsymbol{\Sigma}}

\def\bmu{\boldsymbol{\mu}}

\long\def\symbolfootnote[#1]#2{\begingroup
\def\thefootnote{\fnsymbol{footnote}}\footnote[#1]{#2}\endgroup}

\begin{document}

\title{An extended trivariate vine copula mixed model for  meta-analysis of diagnostic studies in the presence of non-evaluable outcomes}

\author{Aristidis~K.~Nikoloulopoulos \footnote{\href{mailto:a.nikoloulopoulos@uea.ac.uk}{a.nikoloulopoulos@uea.ac.uk},  School of Computing Sciences, University of East Anglia, Norwich NR4 7TJ, U.K.} }
\date{}

\maketitle

\begin{abstract}
\baselineskip=24pt
\noindent
A recent paper proposed an extended trivariate generalized linear mixed model (TGLMM) for synthesis of diagnostic test accuracy studies    
in the presence of non-evaluable  index test results.  Inspired  by the aforementioned model 
we propose an extended trivariate vine copula mixed model that includes the  TGLMM as special case, but can also operate on the original scale of sensitivity, specificity, and disease prevalence. 
The performance of the proposed vine copula mixed model is examined by extensive simulation studies in comparison with the TGLMM. 
Simulation studies showed that the TGLMM   overestimates the meta-analytic estimates of sensitivity,  specificity, and prevalence when the univariate random effects are misspecified. The vine copula mixed model  gives nearly unbiased estimates of test accuracy indices and disease prevalence. 
Our general methodology is  illustrated by meta-analysing  coronary CT angiography studies. 


\end{abstract}

\baselineskip=24pt

\section{Introduction}
Synthesis of diagnostic test accuracy studies is  the most common medical application of multivariate meta-analysis \cite{JacksonRileyWhite2011,
MavridisSalanti13,Ma-etal-2013}.
The purpose of a meta-analysis of diagnostic test accuracy studies is to combine information over different studies, and provide an integrated analysis that will have more statistical power to detect an accurate diagnostic test than an analysis based on a single study.

Diagnostic test accuracy studies observe the result of a gold standard procedure that defines the presence or absence of a disease and the result of a diagnostic test. 
The accuracy of the diagnostic test is commonly measured by a pair of indices such as sensitivity and specificity. Sensitivity is defined as the probability of testing positive given a person being diseased and specificity is defined as the probability of testing negative given a person being non-diseased \cite{Ma-etal-2013}. 
The diagnostic test accuracy studies
typically report the number of true positives (diseased subjects correctly diagnosed), false positives (non-diseased subjects incorrectly diagnosed as diseased), true negatives (non-diseased subjects correctly diagnosed as non-diseased) and false negatives (diseased subjects incorrectly diagnosed as non-diseased). However, diagnostic test outcomes can be non-evaluable
\cite{Begg-etal-1986}. This is the case for   coronary computed tomography (CT) angiography studies which have non-evaluable results of index test in various ways such as when transferring a segment/vessel  to a patient based evaluation \cite{Schuetz-etal-2012}.   

In meta-analysis of diagnostic test accuracy studies, the existence of  non-evaluable subjects  is an important issue that could potentially lead to biased estimates of index test accuracy
\cite{Schuetz-etal-2012,ma-etal-2014}. 
Schuetz et al. \cite{Schuetz-etal-2012} studied different ad-hoc approaches dealing with diagnostic test non-evaluable subjects, such as non-evaluable subjects are excluded from the study, non-evaluable positives (non-evaluable diseased subjects)  are taken as true positives and non-evaluable negatives (non-evaluable non-diseased subjects) are taken as false positives, non-evaluable positives are taken as false negatives and non-evaluable negatives are taken as true negatives, and non-evaluable positives as false negatives and non-evaluable negatives as false positives. 
In all of these approaches,  Schuetz et al. \cite{Schuetz-etal-2012} used the   bivariate generalized linear mixed model (BGLMM) \cite{Chu&Cole2006}, which  assumes independent binomial distributions for the true positives and true negatives, conditional on the latent pair of transformed (via a link function) sensitivity and specificity in each study. They concluded that excluding the index test non-evaluable subjects  leads to overestimation of the meta-analytic estimates of sensitivity and specificity and recommended the  intent-to-diagnose approach by treating non-evaluable positives as false negatives and non-evaluable negatives as false positives.

Ma et al. \cite{ma-etal-2014} proposed a  trivariate generalized linear mixed model (TGLMM)  approach by treating  the non-evaluable subjects as missing data under  a missing at random (MAR) assumption.
The TGLMM was originaly proposed by Chu et al. \cite{chu-etal-2009}  to account for potential correlations among sensitivity,  specificity and disease prevalence as many empirical studies have shown the assumption of independence between the  sensitivity/specificity with disease prevalence for a dichotomous disease status is likely to be violated
\cite{brenner-gefeller-1997,leeflang-etal-2009,Leeflang-etal-2013}. Ma et al. \cite{ma-etal-2014} with extensive simulation studies have shown that (a)  the intent-to-diagnose approach \cite{Schuetz-etal-2012} under-estimates  both meta-analytic estimates of sensitivity and specificity, 
(b)  excluding the index test non-evaluable subjects does not lead to biased estimates of sensitivity and specificity, but leads to biased estimates of prevalence, and (c) the  TGLMM gives nearly unbiased estimates of the meta-analytic estimates of sensitivity, specificity and prevalence.

In this paper, inspired by Ma et al. 
\cite{ma-etal-2014}, we extend the vine copula mixed model  for trivariate meta-analysis of diagnostic test accuracy studies accounting for disease prevalence
\cite{Nikoloulopoulos2015c}  to additionally account for non-evaluable subjects. 
The advantages of this methodology are that  (a)  the extended TGLMM is included  as a special case, (b)  sensitivity, specificity, and prevalence can be modelled in the original scale, and (c) tail dependencies and asymmetries can be provided. 

The remainder of the paper proceeds as follows. Section 2 
introduces the proposed  model for diagnostic test accuracy studies  in the presence of non-evaluable subjects and discusses its relationship with the extended TGLMM. 
Section 3 contains  small-sample  efficiency calculations
to  investigate the effect of misspecifying the random effects distribution on parameter estimates and standard errors and compares the method with the TGLMM approach.  
Section 4 re-evaluates   the meta-analysis of coronary CT angiography studies \cite{Schuetz-etal-2012,ma-etal-2014} using the proposed vine copula mixed model  approach. We conclude with some discussion in Section 5, followed by a brief section with software details.

\section{The vine copula mixed model   in the presence of non-evaluable subjects}
In this section, we extend the trivariate vine copula mixed model \cite{Nikoloulopoulos2015c} to handle non-evaluable results and   discuss its relationship with the extended TGLMM.  

\subsection{Notation}

We first introduce the notation used in this paper. The data are  $y_{ijk},\, i = 1, . . . ,N,\, j=0,1,2,\,k=0,1$, where $i$ is an
index for the individual studies, $j$ is an index for the test outcome (0:negative; 1:positive;  2: non-evaluable) and $k$  is an index for the disease outcome (0: non-diseased; 1: diseased). 
The ``classic" $2\times 2$ table (Table \ref{2times2}) is extended to a $3\times 2$ table (Table \ref{3times2}).  
Each cell in  Tables \ref{2times2} and \ref{3times2} provides  the cell frequency 
corresponding to a combination of index test and disease outcomes in study $i$. Table \ref{3times2} has an additional row that represents the frequencies of non-evaluable outcomes.

\setlength{\tabcolsep}{15pt}
\begin{table}[!h]
\caption{\label{2times2}Data (excluding the non-evaluable outcomes)   from an individual study in a  $2\times 2$  table. }
\centering
\begin{tabular}{ll|cc|c}
\hline
&&\multicolumn{2}{c|}{Disease (by gold standard)}\\  
{Test}&  & $-$ & $+$&Total\\\hline
$-$&  &$y_{i00}$ &$y_{i01}$ & $y_{i0+}$ \\
$+$ &  & $y_{i10}$ & $y_{i11}$&$y_{i1+}$  \\
\hline
Total &  &$y_{i+0}$&$y_{i+1}$&$y_{i++}$\\
\hline
\end{tabular}

\end{table}

\setlength{\tabcolsep}{15pt}
\begin{table}[!h]
\caption{\label{3times2} Data (including the non-evaluable outcomes)  from an individual study in a  $3\times 2$  table. }
\centering
\begin{tabular}{ll|cc|c}
\hline
&&\multicolumn{2}{c|}{Disease (by gold standard)}\\ 
{Test}&  & $-$ & $+$&Total\\\hline
$-$&  & $y_{i00}$ &$y_{i01}$ & $y_{i0+}$ \\
$+$ & & $y_{i10}$&  $y_{i11}$ &$y_{i1+}$  \\
Non-evaluable  &&$y_{i20}$& $y_{i21}$ & $y_{i2+}$\\
\hline
Total &  &$y_{i+0}^*$&$y_{i+1}^*$&$y_{i++}^*$\\

\hline
\end{tabular}

\end{table}

\subsection{The within-study model}

For each study $i$, the within-study model assumes that the 
number  of true negatives $Y_{i00}$,  false negatives $Y_{i01}$, false positives $Y_{i10}$,  true positives $Y_{i11}$, non-evaluable negatives $Y_{i20}$, and    non-evaluable positives $Y_{i21}$
 are multinomially distributed given $\X=\x$, where $\X=(X_1,X_2,X_3,X_4,X_5)$ denotes the transformed (via a link function $l(\cdot)$) latent vector of  sensitivity, specificity, disease prevalence, probability of non-evaluable positives and probability of non-evaluable negatives, viz.
$$
(Y_{i00},Y_{i01},Y_{i10},Y_{i11},Y_{i20},Y_{i21},
)|(X_1=x_1,X_2=x_2,X_3=x_3,X_4=x_4,
X_5=x_5)\sim \mathcal{M}_6\Bigl(y_{i++}^*,\varpibf
\Bigr),\footnote[3]{$\mathcal{M}_T\Bigl(n,\mathbf{p}\Bigr)$  is shorthand notation for the  multinomial distribution; $T$ is the number of cells, $n$ is the number of observations, and  $\mathbf{p}=(p_1,\dots,p_T)$ with $p_1+\ldots+p_T = 1$ is the $T$-dimensional vector of success probabilities.}
$$
where  
\begin{eqnarray}\label{varpibf}
\varpibf&=&(\varpi_{00},\varpi_{01},\varpi_{10},\varpi_{11},\varpi_{20},\varpi_{21})\\
&=&\Bigl(\underbrace{l^{-1}(x_2)\bigl(1-l^{-1}(x_3)\bigr)\bigr(1-l^{-1}(x_5)\bigl)}_{\varpi_{00}},\underbrace{\bigl(1-l^{-1}(x_1)\bigr)l^{-1}(x_3)\bigl(1-l^{-1}(x_4)\bigr)}_{\varpi_{01}},\\
&&\underbrace{\bigl(1-l^{-1}(x_2)\bigr)\bigl(1-l^{-1}(x_3)\bigr)\bigl(1-l^{-1}(x_5)\bigr)}_{\varpi_{10}},\underbrace{l^{-1}(x_1)l^{-1}(x_3)\bigl(1-l^{-1}(x_4)\bigr)}_{\varpi_{11}}, \nonumber\\&&\underbrace{\bigl(1-l^{-1}(x_3)\bigr)l^{-1}(x_5)}_{\varpi_{20}},\underbrace{l^{-1}(x_3)l^{-1}(x_4)}_{\varpi_{21}}
\Bigr)\nonumber
\end{eqnarray}
is derived by Ma et al. \cite{ma-etal-2014} under an MAR assumption.

The multinomial probability mass function (pmf) $$\frac{y_{i++}^*!}{y_{i00}!y_{i01}!y_{i10}!y_{i11}!
y_{i20}!y_{i21}!}\prod_{j=0}^2\prod_{k=0}^1\varpi_{jk}^{y_{ijk}}$$ 
decomposes into a product of independent  binomial pmfs given the random effects, viz.  
$$g\bigl(y_{i11};y_{i+1},l^{-1}(x_1)\bigr) g\bigl(y_{i00};y_{i+0},l^{-1}(x_2)\bigr)g\bigl(y_{i+1}^*;y_{i++}^*,l^{-1}(x_3)\bigr)g\bigl(y_{i21};y_{i+1}^*,l^{-1}(x_4)\bigr)g\bigl(y_{i20};y_{i+0}^*,l^{-1}(x_5)\bigr),$$
where $$g\bigl(y;n,\pi\bigr)=\binom{n}{y}\pi^y(1-\pi)^{n-y},\quad y=0,1,\ldots,n,\quad 0<\pi<1,$$
 is the binomial pmf.
Hence, the within-study model actually  assumes that 
\begin{eqnarray}\label{trivariate-representation}
Y_{i11}|X_{1}=x_1&\sim&\mbox{Binomial}\bigl(y_{i+1},l^{-1}(x_1)\bigr);\nonumber\\
Y_{i00}|X_{2}=x_2&\sim&\mbox{Binomial}\bigl(y_{i+0},l^{-1}(x_2)\bigr);\\
Y_{i+1}^*|X_{3}=x_3&\sim& \mbox{Binomial}\bigl(y_{i++}^*,l^{-1}(x_3)\bigr),\nonumber
\end{eqnarray}
and
\begin{eqnarray*}
Y_{i21}|X_{4}=x_4&\sim& \mbox{Binomial}\bigl(y_{i+1}^*,l^{-1}(x_4)\bigr);\nonumber\\
Y_{i20}|X_{5}=x_5&\sim&  \mbox{Binomial}\bigl(y_{i+0}^*,l^{-1}(x_5)\bigr).\nonumber
\end{eqnarray*}

\subsection{The between-studies model}
Under the MAR assumption, $(X_1,X_2,X_3)$ are  independent of the missing probabilities $(X_4,X_5)$, hence the joint likelihood factors   into two components, one involving only the  transformed sensitivity $x_1$, specificity $x_2$ and disease prevalence $x_3$, and the other involving only the   transformed probabilities of non-evaluable positives $x_4$ and non-evaluable negatives $x_5$. Hence, the methodology of Nikoloulopoulos \cite{Nikoloulopoulos2015c} can be applied to the first likelihood component to infer about the sensitivity, specificity and disease prevalence.

Nikoloulopoulos \cite{Nikoloulopoulos2015c}  proposed a vine copula mixed model as an extension of the TGLMM by rather using a vine copula representation for the random effects distribution of the latent sensitivity,  specificity and disease prevalence. 
The trivariate vine copula
can cover flexible dependence structures through the specification
of $2$ bivariate marginal copulas 
and one
bivariate conditional copula  
that condition on 1 variable \cite{joeetal10}. 
A vine requires a decision on the indexing of variables.  
For a 3-dimensional vine  copula there are $3$ distinct permutations \cite{aasetal09}: $$\{12,13,23|1\}, \qquad \{12,23,13|2\}, \quad \mbox{and} \quad \{13,23,12|3\}.$$ 
To be  concrete in the exposition of the theory,  we use the permutation $\{12,13,23|1\}$; the theory though also apply to the other two permutations.

To this end, the stochastic representation of the between-studies (random effects) model takes the form
\begin{equation}\label{copula-between-norm}
\Bigl(F\bigl(X_1;l(\pi_1),\de_1\bigr),F\bigl(X_2;l(\pi_2),\de_2\bigr)
,F\bigl(X_3;l(\pi_3),\de_3\bigr)\Bigr)\sim C(\cdot;\thbf),
\end{equation}
where 
$C(\cdot;\thbf)$ is a  vine  copula  with dependence parameter vector $\thbf=(\th_{12},\th_{13},\th_{23|1})$ and $F(\cdot;l(\pi),\de)$ is the cdf of the  univariate  distribution of the random effect.  The choices of  $F\bigl(\cdot;l(\pi),\de\bigr)$ and  $l$ are given in Table \ref{choices}. The vine copula density is decomposed in a product of univariate and bivariate copula densities, viz.
\begin{eqnarray*}\label{vine-density}
  f_{123}(x_1,x_2,x_3;\thbf)&=&f_1\bigl(x_1;l(\pi_1),\de_1\bigl)f_2\bigl(x_2;l(\pi_2),\de_2\bigl)f_3\bigl(x_3;l(\pi_3),\de_3\bigl)\,c_{12}\Bigl(F_1\bigl(x_1;l(\pi_1),\de_1\bigl),\nonumber\\&&F_2\bigr(x_2;l(\pi_2),\de_2\bigl);\th_{12}\Bigr)\,c_{13}\Bigl(F_1\bigr(x_1;l(\pi_1),\de_1\bigl),F_3\bigl(x_3;l(\pi_3),\de_3\bigr);\th_{13}\Bigr)\times\nonumber\\&&c_{23|1}\Bigl(F_{2|1}\bigl(x_2|x_1;l(\pi_1),l(\pi_2),\de_1,\de_2\bigr),F_{3|1}\bigl(x_3|x_1;l(\pi_1),l(\pi_3),\de_1,\de_3\bigr);\th_{23|1}\Bigr)\nonumber\\
 &=&f_1\bigl(x_1;l(\pi_1),\de_1\bigl)f_2\bigl(x_2;l(\pi_2),\de_2\bigl)f_3\bigl(x_3;l(\pi_3),\de_3\bigl)c_{123}\Bigl(F_1\bigl(x_1;l(\pi_1),\de_1\bigl),\nonumber\\&&F_2\bigr(x_2;l(\pi_2),\de_2\bigl),F_3\bigl(x_3;l(\pi_3),\de_3\bigr);\thbf\Bigr),
\end{eqnarray*}
where $f_j\bigl(\cdot;l(\pi_j),\de_j\bigl)$ and $F_j\bigl(\cdot;l(\pi_j),\de_j\bigl)$  are the   density  and cdf, respectively, of the random variable $X_j$,  $c_{j_1j_2}(\cdot,\cdot;\th_{j_1j_2})$ and $C_{j_1j_2}(\cdot,\cdot;\th_{j_1j_2})$ are the bivariate copula density and cdf, respectively, for the pair of transformed variables $F_{j_1}(X_{j_1})$ 
and $F_{j_2}(X_{j_2})$,
 and $c_{23|1}(\cdot,\cdot;\th_{23|1})$ is the bivariate copula density for the pair of  transformed variables
$F_{2|1}(X_2|X_1)$ and $F_{3|1}(X_3|X_1)$ where 
$$F_{j_1|j_2}(x_{j_1}|x_{j_2})=\partial C_{j_1j_2}\bigl(F_{j_1}(x_{j_1}),F_{j_2}(x_{j_2})\bigr)/\partial F_{j_2}(x_{j_2})$$ as derived in \cite{joe96}.

\setlength{\tabcolsep}{31pt}
\begin{table}[!h]
\begin{center}
\caption{\label{choices}The choices of the  $F\bigl(\cdot;l(\pi),\de\bigr)$ and  $l$ in the extended trivariate vine copula mixed model.}
\begin{tabular}{cccc}
\toprule $F\bigl(\cdot;l(\pi),\de\bigr)$ & $l$ & $\pi$ & $\de$\\\hline
$N(\mu,\s)$ & logit, probit, cloglog & $l^{-1}(\mu)$&$\s$\\
Beta$(\pi,\gamma)$ & identity & $\pi$ & $\gamma$\\
\bottomrule
\end{tabular}

\end{center}
\end{table}

The copula parameters $\th_{12},\th_{13},\th_{23|1}$ are  parameters of the random effects model, and they are separated from the univariate parameters.  The univariate  parameters  $\pi_1$,  $\pi_2$, and  $\pi_3$ are those of actual interest denoting the meta-analytic parameters of sensitivity, specificity, and disease prevalence, respectively, while the univariate parameters  $\de_1$,  $\de_2$ and $\de_3$ are of secondary interest denoting the variability between-studies for sensitivity, specificity, and disease prevalence, respectively.

\subsection{Likelihood and computational details for maximum likelihood estimation}

For $N$ studies the models in (\ref{trivariate-representation}) and (\ref{copula-between-norm}) together specify a trivariate vine copula mixed  model with joint likelihood
\begin{eqnarray}\label{likelihood}
L(\pi_1,\pi_2,\pi_3,\de_1,\de_2,\de_3,\thbf)&=&
\prod_{i=1}^{N}\int_{0}^{1}\int_{0}^{1}\int_{0}^{1}
g\bigl(y_{i11};y_{i+1},l^{-1}(x_1)\bigr) g\bigl(y_{i00};y_{i+0},l^{-1}(x_2)\bigr)\times \nonumber\\&&g\bigl(y_{i+1}^*;y_{i++}^*,l^{-1}(x_3)\bigr)c_{123}\bigl(u_1,u_2,u_3;\thbf\bigr)du_1du_2du_3,
\end{eqnarray}
where $x_j=F^{-1}\bigl(u_j;l(\pi_j),\de_j\bigr),\, j=1,2,3$.

Estimation of the model parameters  $(\pi_1,\pi_2,\pi_3,\de_1,\de_2,\de_3,\thbf)$  can be approached by the standard maximum likelihood (ML) method, by maximizing the logarithm of the joint likelihood in (\ref{likelihood}). 
The estimated parameters can be obtained by 
using a quasi-Newton \cite{nash90} method applied to the logarithm of the joint likelihood.  
This numerical  method requires only the objective
function, i.e.,  the logarithm of the joint likelihood, while the gradients
are computed numerically and the Hessian matrix of the second
order derivatives is updated in each iteration. The standard errors (SE) of the ML estimates can be also obtained via the gradients and the Hessian computed numerically during the maximization process.

For the vine copula mixed model numerical evaluation of the joint pmf can be achieved with the following steps:

\begin{enumerate}
\itemsep=10pt
\item Calculate Gauss-Legendre \cite{Stroud&Secrest1966}  quadrature points $\{u_q: q=1,\ldots,n_q\}$ 
and weights $\{w_q: q=1,\ldots,n_q\}$ in terms of standard uniform.
\item Convert from independent uniform random variables $\{u_{q_1}: q_1=1,\ldots,n_q\}$,  $\{u_{q_2}: q_2=1,\ldots,n_q\}$, and $\{u_{q_3}: q_3=1,\ldots,n_q\}$ to   dependent uniform random variables $v_{q_1},v_{q_2|q_1}$, and $v_{q_2q_3|q_1}$ that have a vine distribution $C(\cdot;\thbf)$ \cite{Nikoloulopoulos2015c}:
\begin{algorithmic}[1]
\State Set  $v_{q_1}=u_{q_1}$   
\State $v_{q_2|q_1}=C^{-1}_{12}(u_{q_2}|u_{q_1};\th_{12})$
\State $t_1=C^{-1}_{23|1}(u_{q_3}|u_{q_2};$ $\th_{23|1})$
\State $v_{q_2q_3|q_1}=C^{-1}_{13}\Bigl(t_1|u_{q_1};\th_{13}\Bigr)$,
\end{algorithmic}
where  $C(v|u;\th)$ and  $C^{-1}(v|u;\th)$ are  the conditional copula cdf 
and its inverse. 

\item Numerically evaluate the joint pmf
\begin{align*}
&\int_{0}^{1}\int_{0}^{1}\int_{0}^{1}
g\left(y_{i11};y_{i+1},l^{-1}\Bigl(F^{-1}\bigl(u_1;l(\pi_1),\de_1\bigr)\Bigr)\right) g\left(y_{i00};y_{i+0},l^{-1}\Bigl(F^{-1}\bigl(u_2;l(\pi_2),\de_2\bigr)\Bigr)\right)\\&g\left(y_{i+1}^*;y_{i++}^*,l^{-1}\Bigl(F^{-1}\bigl(u_3;l(\pi_3),\de_3\bigr)\Bigr)\right)c_{123}(u_1,u_2,u_3;\thbf)du_1du_2du_3
\end{align*}
in a triple sum:
\begin{align*}
&\sum_{q_1=1}^{n_q}\sum_{q_2=1}^{n_q}\sum_{q_3=1}^{n_q}g\left(y_{i11};y_{i+1},l^{-1}\Bigl(F^{-1}\bigl(v_{q_1};l(\pi_1),\de_1\bigr)\Bigr)\right) g\left(y_{i00};y_{i+0},l^{-1}\Bigl(F^{-1}\bigl(v_{q_2|q_1};l(\pi_2),\de_2\bigr)\Bigr)\right)\\&g\left(y_{i+1}^*;y_{i++}^*,l^{-1}\Bigl(F^{-1}\bigl(v_{q_2q_3|q_1};l(\pi_3),\de_3\bigr)\Bigr)\right).
\end{align*}

\end{enumerate}

With Gauss-Legendre quadrature, the same nodes and weights
are used for different functions;
this helps in yielding smooth numerical derivatives for numerical optimization via quasi-Newton.

\subsection{\label{special}Relationship with  the TGLMM}
In this subsection, we show what happens when all the bivariate copulas are bivariate normal (BVN)  and the univariate distribution of the random effects is the  $N(\mu,\s)$ distribution.  
One can easily deduce that the within-study model in (\ref{trivariate-representation}) is the same as in the TGLMM. 

Furthermore, when the three bivariate copulas are BVN copulas with copula (correlation) parameters  $\rho_{12},\rho_{13},\rho_{23|1}$, 
the resulting distribution is the trivariate normal (TVN)  with mean vector $\bmu=\bigl(l(\pi_1),l(\pi_2),l(\pi_3)\bigr)^\top$ and variance covariance matrix
$\Sbf=\begin{pmatrix}
\sigma_1^2 &\rho_{12}\sigma_1\s_2 &\rho_{13}\sigma_1\s_3\\
\rho_{12}\sigma_1\sigma_2 & \sigma_2^2&\rho_{23}\sigma_2\s_3&\\
\rho_{13}\sigma_1\sigma_3 &\rho_{23}\sigma_2\s_3&\s_3^2\\
\end{pmatrix},$ where
$\rho_{23}=\rho_{23|1}\sqrt{1-\rho_{12}^2}\sqrt{1-\rho_{13}^2}\rho_{12}\rho_{13}$.
Therefore,  the between-studies model in (\ref{copula-between-norm}) 
 assumes
that $\X=(X_1,X_2,X_3)$ is TVN distributed, i.e., $
\X
\sim
\mbox{TVN}
\bigl(\bmu,\Sbf\bigr).
$

With some calculus it can be shown that the joint likelihood in (\ref{likelihood})
becomes
\begin{eqnarray*}\label{likelihood}
L(\pi_1,\pi_2,\pi_3,\s_1,\s_2,\s_3,\thbf)&=&\
\prod_{i=1}^{N}\int_{-\infty}^{\infty}\int_{-\infty}^{\infty}\int_{-\infty}^{\infty}
g\bigl(y_{i11};y_{i+1},l^{-1}(x_1)\bigr) g\bigl(y_{i00};y_{i+0},l^{-1}(x_2)\bigr)\times\\
&&g\bigl(y_{i+1}^*;y_{i++}^*,l^{-1}(x_3)\bigr)
\phi_{123}(x_1,x_2,x_3;\bmu,\Sbf)dx_1dx_2dx_3,
\end{eqnarray*}
 where 
 $\phi_{123}(\cdot;\bmu,\Sbf)$ is the TVN density with mean vector $\bmu$  and variance covariance matrix $\Sbf$. Hence,  this model is the same as the extended TGLMM in \cite{ma-etal-2014}.

\section{Small-sample efficiency--misspecification of the random effects distribution}
An extensive simulation study is conducted  
(a) to gauge the small-sample efficiency of the ML 
method, and 
(b) to investigate in detail 
the  misspecification of the parametric margin or  family of copulas of the random effects distribution.     We also include comparisons with the TGLMM
\cite{ma-etal-2014}, that is a vine copula mixed model composed of BVN copulas and normal margins as shown in Subsection \ref{special},  as the current state of the art of the various  meta-analytic approaches to handle non-evaluable results.  We don't include either the intent-to-diagnose approach or  the BGLMM that excludes the non-evaluable subjects   in our simulation study as Ma et al. \cite{ma-etal-2014} has already established that these methods produce biased estimates in the presence of index test non-evaluable subjects. 

In this simulation study we follow the configurations in  Ma et al. \cite{ma-etal-2014}. 
We conduct simulation studies under three missing  
scenarios: 
\begin{itemize}
\itemsep=10pt
\item the probabilities for non-evaluable diseased and non-diseased subjects are the same, i.e.,  $l^{-1}(x_4)=l^{-1}(x_5)=0.1$;
\item  the probability for non-evaluable diseased subjects is smaller than the probability for non-evaluable  non-diseased subjects, i.e.,  $l^{-1}(x_4)=0.1<l^{-1}(x_5)=0.2$;
\item the probability for non-evaluable non-diseased subjects is smaller than the probability for non-evaluable  diseased subjects, i.e., $l^{-1}(x_4)=0.2>l^{-1}(x_5)=0.1$. 
\end{itemize}
All three scenarios satisfy the MAR assumption, and the first scenario also satisfies the missing completely at random  assumption \cite{Little&Rubin2002}. 

True sensitivity $\pi_1$ and specificity $\pi_2$ are 0.7 and 0.9, disease prevalence $\pi_3$ is 0.25 and the variability parameters are $\s_1=\s_2=\s_3=1$ or $\g_1=\g_2=\g_3=0.1$ for normal or beta margin, respectively. A moderate negative Kendall's tau association  of $\tau_{12}=-0.5$ is assumed between $X_1$ and $X_2$, a moderate positive Kendall's tau association   of $\tau_{13}=0.5$ is assumed between $X_1$ and $X_3$, and a moderate negative Kendall's tau association  of $\tau_{23|1}=-0.5$ is assumed between $X_2$ and $X_3$ given $X_1$.   Under each setting, 10,000 meta-analysis data sets are simulated with $N=30$ studies in each data set.
The simulation process is as
below:

\noindent For $i=1,\ldots,N$:
\begin{enumerate}
\itemsep=10pt
\item Simulate $(u_1,u_2,u_3)$ from a C-vine $C(\cdot;\tau_{12},\tau_{13},\tau_{23|1})$ \cite{joe2010a}.  We convert from $\tau$'s to the BVN, Frank and   (rotated)  Clayton copula parameters $\theta$'s via the relations 
\begin{equation}\label{tauBVN}
\tau=\frac{2}{\pi}\arcsin(\th),
\end{equation}
\begin{equation}\label{tauFrank}
\tau=\left\{\begin{array}{ccc}
1-4\theta^{-1}-4\theta^{-2}\int_\theta^0\frac{t}{e^t-1}dt &,& \th<0\\
1-4\theta^{-1}+4\theta^{-2}\int^\theta_0\frac{t}{e^t-1}dt &,& \th>0\\
\end{array}\right.,
\end{equation}
and 
\begin{equation}\label{tauCln}
\tau=\left\{\begin{array}{rcl}
\th/(\th+2)&,& \mbox{by 0$^\circ$ or 180$^\circ$ }\\
-\th/(\th+2)&,& \mbox{by 90$^\circ$ or 270$^\circ$}\\
\end{array}\right.,
\end{equation}
in \cite{HultLindskog02}, \cite{genest87}, and 
\cite{genest&mackay86}, respectively.  

\item Convert to beta  or normal realizations via $x_j=l^{-1}\Bigl(F_j^{-1}\bigl(u_j,l(\pi_j),\de_j\bigr)\Bigr)$ for $j=1,2,3$.

\item Simulate the study size $n$ from a shifted gamma distribution \cite{paul-etal-2010}, i.e., $n\sim \mbox{sGamma}(\a=1.2,\b=0.01,\mbox{lag}=30)$ and round off to the nearest integer. 
\item Generate  $(y_{i00},y_{i01},y_{i10},y_{i11},
y_{i20},y_{i21})$   from  $\mathcal{M}_6(n,\varpibf)$;
see (\ref{varpibf}) for the elements    of the probability vector  $\varpibf$. 
\end{enumerate}

 \setlength{\tabcolsep}{4pt}

\begin{table}[!h]
  \centering
 \caption{\label{sim-norm-01.02}Biases,  root mean square errors (RMSE) and standard deviations (SD), along with the square root of the average theoretical variances ($\sqrt{\bar V}$), scaled by 100, for the MLEs  under different copula choices and margins from  $10^4$ small sample of sizes $N = 30$ simulations ($n_q=15$) from the trivariate vine copula mixed model with  Clayton  copulas  rotated by $90^\circ$  for both the $C_{12}(;\tau_{12})$ and $C_{13}(;\tau_{23|1})$ copulas, the Clayton copula for the $C_{13}(;\tau_{13})$ copula and normal margins. The missing probability of diseased group is smaller than non-diseased group, i.e.,  $v_4=0.1<v_5=0.2$.}
    \begin{tabular}{lllccccccccc}
    \toprule
          & margin & copula & $\pi_1$ & $\pi_2$ & $\pi_3$ & $\s_1$ & $\s_2$ & $\s_3$ & $\tau_{12}$ & $\tau_{13}$ & $\tau_{23|1}$ \\
    \midrule
   \rowcolor{gray}     Bias &$^\dag$ normal & BVN   & 0.19  & -0.12 & 0.11  & -7.96 & -2.36 & -1.90 & -4.87 & -3.33 & 16.76 \\
          &       & Frank & 0.22  & -0.16 & -0.27 & -7.44 & -1.47 & -2.31 & -6.24 & -1.14 & 15.21 \\
    \rowcolor{gray}         & $^\S$      & Cln$\{0^\circ,90^\circ\}$ & 0.32  & -0.04 & 0.88  & -6.96 & 1.92  & -3.97 & -3.73 & -6.85 & 20.71 \\
          &       & Cln$\{0^\circ,270^\circ\}$ & 1.27  & -0.22 & 0.01  & -9.14 & -1.80 & 1.15  & 4.94  & -17.02 & 16.67 \\
    \rowcolor{gray}         & beta  & BVN   & -2.07 & -3.57 & 4.04  & -     & -     & -     & -3.93 & -4.68 & 18.46 \\
          &       & Frank & -1.76 & -3.56 & 3.55  & -     & -     & -     & -5.90 & -1.81 & 17.34 \\
   \rowcolor{gray}          &      & Cln$\{0^\circ,90^\circ\}$ & -1.82 & -3.71 & 4.48  & -     & -     & -     & -2.78 & -9.76 & 19.03 \\
          &       & Cln$\{0^\circ,270^\circ\}$ & -1.14 & -3.72 & 4.15  & -     & -     & -     & 5.39  & -17.35 & 19.60 \\\hline
  \rowcolor{gray}      SD &$^\dag$ normal & BVN   & 4.37  & 1.86  & 3.55  & 17.84 & 16.56 & 14.73 & 15.38 & 13.20 & 23.98 \\
          &       & Frank & 4.61  & 1.93  & 3.60  & 18.59 & 17.28 & 14.81 & 16.66 & 14.04 & 25.02 \\
  \rowcolor{gray}           & $^\S$      & Cln$\{0^\circ,90^\circ\}$ & 4.53  & 1.91  & 3.79  & 19.66 & 18.88 & 15.64 & 14.28 & 17.85 & 24.96 \\
          &       & Cln$\{0^\circ,270^\circ\}$ & 4.28  & 2.00  & 3.71  & 18.96 & 18.48 & 17.35 & 27.31 & 17.87 & 22.77 \\
    \rowcolor{gray}         & beta  & BVN   & 3.95  & 2.27  & 3.37  & 3.97  & 3.03  & 3.35  & 15.08 & 13.19 & 23.61 \\
          &       & Frank & 4.13  & 2.34  & 3.38  & 4.14  & 3.07  & 3.34  & 16.45 & 14.03 & 25.55 \\
    \rowcolor{gray}         &     & Cln$\{0^\circ,90^\circ\}$ & 4.10  & 2.43  & 3.67  & 4.35  & 3.69  & 3.53  & 14.52 & 17.58 & 26.02 \\
          &       & Cln$\{0^\circ,270^\circ\}$ & 3.91  & 2.41  & 3.58  & 4.21  & 3.19  & 4.13  & 26.97 & 18.78 & 22.80 \\\hline
  \rowcolor{gray}      $\sqrt{\bar V}$ &$^\dag$ normal & BVN   & 3.95  & 1.77  & 3.36  & 16.76 & 15.84 & 13.52 & 12.73 & 11.05 & 18.65 \\
          &       & Frank & 3.90  & 1.74  & 3.20  & 17.04 & 16.13 & 13.19 & 12.38 & 10.96 & 17.46 \\
   \rowcolor{gray}          & $^\S$      & Cln$\{0^\circ,90^\circ\}$ & 3.73  & 1.69  & 2.83  & 16.04 & 14.72 & 10.77 & 11.20 & 8.37  & 13.52 \\
          &       & Cln$\{0^\circ,270^\circ\}$ & 3.42  & 1.65  & 3.02  & 15.86 & 14.61 & 13.07 & 11.66 & 7.72  & 10.44 \\
    \rowcolor{gray}         & beta  & BVN   & 3.54  & 2.00  & 3.04  & 3.78  & 2.54  & 3.02  & 12.79 & 11.22 & 18.70 \\
          &       & Frank & 3.51  & 1.94  & 2.89  & 3.87  & 2.54  & 2.88  & 12.68 & 11.04 & 17.78 \\
    \rowcolor{gray}         &     & Cln$\{0^\circ,90^\circ\}$ & 3.41  & 1.83  & 2.48  & 3.58  & 2.46  & 2.30  & 11.57 & 8.79  & 13.99 \\
          &       & Cln$\{0^\circ,270^\circ\}$ & 3.13  & 1.83  & 2.70  & 3.53  & 2.31  & 2.82  & 12.26 & 8.12  & 10.35 \\\hline
    \rowcolor{gray}    RMSE &$^\dag$ normal & BVN   & 4.37  & 1.86  & 3.55  & 19.54 & 16.72 & 14.85 & 16.13 & 13.61 & 29.26 \\
          &       & Frank & 4.62  & 1.93  & 3.61  & 20.02 & 17.34 & 14.99 & 17.79 & 14.09 & 29.28 \\
    \rowcolor{gray}         & $^\S$      & Cln$\{0^\circ,90^\circ\}$ & 4.54  & 1.91  & 3.89  & 20.86 & 18.98 & 16.14 & 14.76 & 19.11 & 32.43 \\
          &       & Cln$\{0^\circ,270^\circ\}$ & 4.46  & 2.01  & 3.71  & 21.05 & 18.56 & 17.39 & 27.75 & 24.67 & 28.22 \\
   \rowcolor{gray}          & beta  & BVN   & 4.46  & 4.24  & 5.26  & -     & -     & -     & 15.58 & 14.00 & 29.97 \\
          &       & Frank & 4.49  & 4.26  & 4.90  & -     & -     & -     & 17.48 & 14.15 & 30.88 \\
     \rowcolor{gray}        &       & Cln$\{0^\circ,90^\circ\}$ & 4.49  & 4.44  & 5.79  & -     & -     & -     & 14.78 & 20.11 & 32.24 \\
          &       & Cln$\{0^\circ,270^\circ\}$ & 4.07  & 4.43  & 5.48  & -     & -     & -     & 27.50 & 25.57 & 30.06 \\
    \bottomrule
    \end{tabular}%
     \vspace{-1ex}
\begin{flushleft}
\begin{footnotesize}
$^\S$: True model;
$^\dag$: The resulting model is the same as the TGLMM; Cln\{$\omega_1^\circ,\omega_2^\circ$\}: The $C_{13}(\cdot;\tau_{13})$ and \{$C_{12}(\cdot;\tau_{12}), C_{23|1}(\cdot;\tau_{23|1})$\} pair copulas are Clayton rotated by $\omega_1$ and $\omega_2$ degrees, respectively.
\end{footnotesize}  
\end{flushleft}
  \label{tab:addlabel}%
\end{table}

\begin{table}[!h]
  \centering
 \caption{\label{sim-beta-01.02}Biases,  root mean square errors (RMSE) and standard deviations (SD), along with the square root of the average theoretical variances ($\sqrt{\bar V}$), scaled by 100, for the MLEs  under different copula choices and margins from  $10^4$ small sample of sizes $N = 30$ simulations ($n_q=15$) from the trivariate vine copula mixed model with  Clayton  copulas  rotated by $90^\circ$  for both the $C_{12}(;\tau_{12})$ and $C_{13}(;\tau_{23|1})$ copulas, the Clayton copula for the $C_{13}(;\tau_{13})$ copula and beta margins.  The missing probability of diseased group is smaller than non-diseased group, i.e.,  $v_4=0.1<v_5=0.2$.}
    \begin{tabular}{lllccccccccc}
    \toprule
          & margin & copula & $\pi_1$ & $\pi_2$ & $\pi_3$ & $\g_1$ & $\g_2$ & $\g_3$ & $\tau_{12}$ & $\tau_{13}$ & $\tau_{23|1}$ \\
    \midrule
   \rowcolor{gray}     Bias &$^\dag$ normal & BVN   & 2.25  & 3.59  & -2.79 & -     & -     & -     & -6.31 & -2.51 & 14.22 \\
          &       & Frank & 2.10  & 3.53  & -2.92 & -     & -     & -     & -7.57 & -0.71 & 11.71 \\
     \rowcolor{gray}        &      & Cln$\{0^\circ,90^\circ\}$ & 2.16  & 3.68  & -2.54 & -     & -     & -     & -4.18 & -2.36 & 15.41 \\
          &       & Cln$\{0^\circ,270^\circ\}$ & 2.96  & 3.51  & -2.80 & -     & -     & -     & 0.28  & -12.14 & 10.29 \\
     \rowcolor{gray}        & beta  & BVN   & 0.64  & -0.02 & 0.10  & -1.51 & -0.37 & -0.57 & -6.72 & -3.62 & 17.75 \\
          &       & Frank & 0.71  & -0.11 & -0.13 & -1.50 & -0.19 & -0.68 & -8.26 & -1.04 & 16.53 \\
     \rowcolor{gray}        & $^\S$      & Cln$\{0^\circ,90^\circ\}$ & 0.71  & -0.01 & 0.19  & -1.54 & -0.03 & -1.03 & -4.13 & -5.44 & 14.81 \\
          &       & Cln$\{0^\circ,270^\circ\}$ & 1.34  & -0.22 & 0.20  & -1.91 & -0.23 & -0.17 & -0.65 & -12.45 & 14.28 \\\hline
    \rowcolor{gray}    SD &$^\dag$ normal & BVN   & 3.50  & 1.60  & 2.77  & 16.20 & 21.16 & 13.06 & 17.23 & 15.58 & 30.77 \\
          &       & Frank & 3.68  & 1.66  & 2.80  & 16.59 & 22.27 & 13.15 & 18.74 & 16.76 & 31.27 \\
    \rowcolor{gray}         &     & Cln$\{0^\circ,90^\circ\}$ & 3.66  & 1.65  & 2.91  & 17.36 & 22.00 & 13.87 & 14.82 & 20.92 & 28.61 \\
          &       & Cln$\{0^\circ,270^\circ\}$ & 3.44  & 1.71  & 2.85  & 17.04 & 23.62 & 14.94 & 28.32 & 20.78 & 29.77 \\
     \rowcolor{gray}        & beta  & BVN   & 3.25  & 1.86  & 2.60  & 3.16  & 2.97  & 2.42  & 17.38 & 15.53 & 31.40 \\
          &       & Frank & 3.38  & 1.96  & 2.61  & 3.23  & 3.13  & 2.39  & 19.16 & 16.66 & 32.36 \\
     \rowcolor{gray}        & $^\S$      & Cln$\{0^\circ,90^\circ\}$ & 3.38  & 1.94  & 2.74  & 3.35  & 3.26  & 2.44  & 15.01 & 20.93 & 29.30 \\
          &       & Cln$\{0^\circ,270^\circ\}$ & 3.21  & 2.01  & 2.70  & 3.27  & 3.26  & 2.81  & 29.16 & 22.00 & 29.70 \\\hline
     \rowcolor{gray}   $\sqrt{\bar V}$ &$^\dag$ normal & BVN   & 3.25  & 1.43  & 2.63  & 14.90 & 20.36 & 11.73 & 14.34 & 13.37 & 24.48 \\
          &       & Frank & 3.21  & 1.43  & 2.53  & 14.87 & 20.93 & 11.60 & 13.08 & 13.11 & 21.31 \\
   \rowcolor{gray}          &       & Cln$\{0^\circ,90^\circ\}$ & 3.14  & 1.35  & 2.34  & 14.45 & 18.53 & 10.01 & 11.83 & 10.51 & 16.45 \\
          &       & Cln$\{0^\circ,270^\circ\}$ & 2.89  & 1.37  & 2.51  & 14.12 & 19.35 & 11.70 & 11.18 & 10.26 & 13.86 \\
      \rowcolor{gray}       & beta  & BVN   & 3.05  & 1.80  & 2.48  & 3.01  & 2.84  & 2.30  & 14.61 & 13.44 & 24.71 \\
          &       & Frank & 3.04  & 1.79  & 2.40  & 3.02  & 2.93  & 2.24  & 13.96 & 13.34 & 22.40 \\
      \rowcolor{gray}       & $^\S$      & Cln$\{0^\circ,90^\circ\}$ & 3.00  & 1.63  & 2.16  & 2.88  & 2.60  & 1.88  & 12.54 & 11.10 & 17.33 \\
          &       & Cln$\{0^\circ,270^\circ\}$ & 2.78  & 1.71  & 2.40  & 2.80  & 2.73  & 2.29  & 12.37 & 10.98 & 14.92 \\\hline
    \rowcolor{gray}    RMSE &$^\dag$ normal & BVN   & 4.16  & 3.93  & 3.93  & -     & -     & -     & 18.35 & 15.78 & 33.90 \\
          &       & Frank & 4.24  & 3.90  & 4.05  & -     & -     & -     & 20.21 & 16.77 & 33.39 \\
   \rowcolor{gray}          &     & Cln$\{0^\circ,90^\circ\}$ & 4.25  & 4.03  & 3.87  & -     & -     & -     & 15.39 & 21.05 & 32.50 \\
          &       & Cln$\{0^\circ,270^\circ\}$ & 4.53  & 3.90  & 4.00  & -     & -     & -     & 28.32 & 24.07 & 31.50 \\
       \rowcolor{gray}      & beta  & BVN   & 3.31  & 1.86  & 2.60  & 3.50  & 3.00  & 2.49  & 18.63 & 15.94 & 36.07 \\
          &       & Frank & 3.46  & 1.96  & 2.62  & 3.56  & 3.13  & 2.49  & 20.87 & 16.69 & 36.33 \\
     \rowcolor{gray}        & $^\S$      & Cln$\{0^\circ,90^\circ\}$ & 3.45  & 1.94  & 2.75  & 3.69  & 3.26  & 2.65  & 15.56 & 21.62 & 32.83 \\
          &       & Cln$\{0^\circ,270^\circ\}$ & 3.48  & 2.03  & 2.71  & 3.79  & 3.27  & 2.82  & 29.17 & 25.27 & 32.95 \\
    \bottomrule
    \end{tabular}%
     \vspace{-1ex}
\begin{flushleft}
\begin{footnotesize}
$^\S$: True model;
$^\dag$: The resulting model is the same as the TGLMM; Cln\{$\omega_1^\circ,\omega_2^\circ$\}: The $C_{13}(\cdot;\tau_{13})$ and \{$C_{12}(\cdot;\tau_{12}), C_{23|1}(\cdot;\tau_{23|1})$\} pair copulas are Clayton rotated by $\omega_1$ and $\omega_2$ degrees, respectively.
\end{footnotesize}  
\end{flushleft}
  \label{tab:addlabel}%
\end{table}

From the simulation results it is revealed that the   MLEs   are not  affected by different missingness scenarios. Hence  we provide here the simulation results for one missingness scenario (Tables  \ref{sim-norm-01.02} and \ref{sim-beta-01.02}). The results for  the other two missingness scenarios are provided in the tables of the Supplementary Material. The tables
 contain the  resultant biases,  standard deviations (SDs),  average theoretical variances $\sqrt{\bar V}$, and root mean square errors (RMSEs), scaled by 100, for the maximum likelihood estimates (MLEs)  under different copula choices and margins    under different copula and marginal choices from  the vine copula mixed model with  normal and beta margins, respectively.  
The theoretical variances of the MLEs are obtained via the gradients and the Hessian that were computed numerically during the maximization process.  The true (simulated) copula distributions are the Clayton  copulas  rotated by $90^\circ$  for both the $C_{12}(\cdot;\tau_{12})$ and $C_{13}(\cdot;\tau_{23|1})$ copulas and the Clayton copula for the $C_{13}(\cdot;\tau_{13})$ copula.

Conclusions from the values in the tables are the following:

\begin{itemize}

\item ML   with  the true vine copula mixed  model is highly efficient according to the simulated biases,  SDs and RMSEs.

\item The ML estimates of the univariate parameters and their SDs are robust under copula misspecification,  but are not robust to margin misspecification.

\item The  ML  estimates of $\tau$'s  and their SDs are robust to margin misspecification, as the copula remains invariant under any series of strictly increasing transformations of the components of the random vector.

 \end{itemize}

  These results are in line with our previous studies   \cite{Nikoloulopoulos2015b,Nikoloulopoulos2015c,Nikoloulopoulos-2016-SMMR,Nikoloulopoulos2018-AStA,nikoloulopoulos-2018-smmr}. The estimation of the univariate meta-analytic parameters is a univariate inference,  and hence it is the univariate marginal distribution that  matters and not the type of the copula. The extended TGLMM \cite{ma-etal-2014} assumes  normal margins; this  is too restrictive and, as shown in Table \ref{sim-beta-01.02} and Supplementary Tables 3 and 4, leads to overestimation of the meta-analytic univariate parameters when the true univariate distribution  of the latent sensitivity, specificity, and disease prevalence  is  beta.

\section{\label{sec-app}Re-evaluation of the meta-analysis of coronary CT angiography studies}

We illustrate the use of the vine copula mixed model for the meta-analysis of diagnostic accuracy studies in the presence of non-evaluable subjects  by re-analysing the data on 26 studies from a  systematic review for diagnostic accuracy studies of coronary CT angiography    \cite{Schuetz-etal-2012,ma-etal-2014}.
 
We fit the vine copula mixed model for all different permutations, 
choices of parametric families of copulas and margins. To make it easier
to compare strengths of dependence, we convert from $\tau$ to the BVN, Frank  and   (rotated)  Clayton copula parameter $\theta$ via the relations in (\ref{tauBVN}), (\ref{tauFrank}), and (\ref{tauCln}).  Since the number of parameters is the same between the models, we use the maximized log-likelihood that corresponds to the estimates as a rough diagnostic measure for goodness of fit between the models.

In Table \ref{app-table} we present the results from the first permutation, as a different indexing didn't lead to any significant differences due to the small sample size. This is consistent with our previous study on trivariate vine copula mixed models  \cite{Nikoloulopoulos2015c}. The log-likelihoods showed that a vine copula mixed model with the Clayton copula for the $C_{12}(\cdot;\tau_{12})$ copula and  the Clayton  copula  rotated by $90^\circ$  for both the $C_{13}(\cdot;\tau_{13})$ and $C_{23|1}(\cdot;\tau_{23|1})$ copulas and beta margins provides the best fit (Table \ref{app-table}).
It is also revealed that a vine copula mixed model with the sensitivity, specificity, and prevalence on the original scale  provides better fit than the TGLMM, which models the sensitivity, specificity and prevalence on a transformed scale.

\setlength{\tabcolsep}{9pt}
\begin{table}[!h]
  \centering
  \caption{\label{app-table}Maximised log-likelihoods, ML estimates and standard errors (SE) of the  trivariate  vine copula mixed models for  diagnostic accuracy studies of coronary CT angiography.}
    \begin{tabular}{ccccccccccccc}
    \toprule\multicolumn{13}{c}{Normal margins}  \\\hline       
          &       & \multicolumn{2}{c}{BVN $^\dag$} &       & \multicolumn{2}{c}{Cln$\{0^\circ,90^\circ\}$} &       & \multicolumn{2}{c}{Cln$\{0^\circ,270^\circ\}$} &       & \multicolumn{2}{c}{Frank} \\
    \cmidrule{3-4}  \cmidrule{6-7}\cmidrule{9-10} \cmidrule{12-13}
          &       & Est.  & SE    &       & Est.  & SE    &       & Est.  & SE    &       & Est.  & SE \\\hline
    $\pi_1$ &       & 0.982 & 0.006 &       & 0.982 & 0.006 &       & 0.982 & 0.006 &       & 0.980 & 0.005 \\
    $\pi_2$ &       & 0.890 & 0.021 &       & 0.892 & 0.021 &       & 0.891 & 0.021 &       & 0.885 & 0.022 \\
    $\pi_3$ &       & 0.481 & 0.040 &       & 0.482 & 0.039 &       & 0.484 & 0.039 &       & 0.478 & 0.039 \\
    $\s_1$ &       & 0.687 & 0.343 &       & 0.670 & 0.347 &       & 0.684 & 0.328 &       & 0.478 & 0.291 \\
    $\s_2$ &       & 0.866 & 0.200 &       & 0.863 & 0.198 &       & 0.843 & 0.189 &       & 0.878 & 0.190 \\
    $\s_3$ &       & 0.790 & 0.115 &       & 0.781 & 0.094 &       & 0.808 & 0.104 &       & 0.753 & 0.118 \\
    $\tau_{12}$ &       & 0.539 & 0.374 &       & 0.391 & 0.375 &       & 0.439 & 0.364 &       & 0.815 & 0.203 \\
    $\tau_{13}$ &       & -0.110 & 0.227 &       & 0.018 & 0.346 &       & -0.058 & 0.108 &       & -0.026 & 0.260 \\
    $\tau_{23|1}$ &       & -0.231 & 0.312 &       & -0.320 & 0.281 &       & -0.040 & 0.128 &       & -0.911 & 0.132 \\\hline
    $\log L$ &       & \multicolumn{2}{c}{$-194.9$} &       & \multicolumn{2}{c}{$-194.3$} &       & \multicolumn{2}{c}{$-195.4$} &       & \multicolumn{2}{c}{$-194.4$} \\\hline\multicolumn{13}{c}{Beta margins}  \\\hline   
          &       & \multicolumn{2}{c}{BVN} &       & \multicolumn{2}{c}{Cln$\{0^\circ,90^\circ\}$ $^\S$} &       & \multicolumn{2}{c}{Cln$\{0^\circ,270^\circ\}$} &       & \multicolumn{2}{c}{Frank} \\   \cmidrule{3-4}  \cmidrule{6-7}\cmidrule{9-10} \cmidrule{12-13}
          &       & Est.  & SE    &       & Est.  & SE    &       & Est.  & SE    &       & Est.  & SE \\\hline
    $\pi_1$ &       & 0.978 & 0.006 &       & 0.977 & 0.006 &       & 0.978 & 0.005 &       & 0.977 & 0.005 \\
    $\pi_2$ &       & 0.864 & 0.022 &       & 0.865 & 0.022 &       & 0.865 & 0.021 &       & 0.856 & 0.023 \\
    $\pi_3$ &       & 0.484 & 0.034 &       & 0.483 & 0.032 &       & 0.487 & 0.032 &       & 0.480 & 0.034 \\
    $\g_1$ &       & 0.010 & 0.011 &       & 0.009 & 0.011 &       & 0.009 & 0.011 &       & 0.005 & 0.006 \\
    $\g_2$ &       & 0.076 & 0.031 &       & 0.075 & 0.031 &       & 0.073 & 0.029 &       & 0.081 & 0.031 \\
    $\g_3$ &       & 0.118 & 0.027 &       & 0.115 & 0.022 &       & 0.123 & 0.025 &       & 0.110 & 0.028 \\
    $\tau_{12}$ &       & 0.519 & 0.367 &       & 0.407 & 0.456 &       & 0.465 & 0.490 &       & 0.797 & 0.205 \\
    $\tau_{13}$ &       & -0.105 & 0.225 &       & 0.033 & 0.272 &       & -0.057 & 0.107 &       & -0.016 & 0.252 \\
    $\tau_{23|1}$ &       & -0.241 & 0.282 &       & -0.345 & 0.234 &       & -0.040 & 0.124 &       & -0.911 & 0.104 \\\hline
    $\log L$ &       & \multicolumn{2}{c}{$-194.5$} &       & \multicolumn{2}{c}{$-193.9$} &       & \multicolumn{2}{c}{$-195.2$} &       & \multicolumn{2}{c}{$-194.0$} \\
    \bottomrule
    \end{tabular}%
   \begin{flushleft}
\begin{footnotesize}
$^\S$: Best fit; $^\dag$: The resulting model is the same as the TGLMM; Cln\{$\omega_1^\circ,\omega_2^\circ$\}: The $C_{12}(\cdot;\tau_{12})$ and \{$C_{13}(\cdot;\tau_{13})$,  $C_{23|1}(\cdot;\tau_{23|1})$\} pair copulas are Clayton rotated by $\omega_1$ and $\omega_2$ degrees, respectively. 
\end{footnotesize}  
\end{flushleft} 
\end{table}

\begin{figure}[!h]
\caption{\label{SROCs}Contour plots (predictive region)  and quantile  regression curves  from the best fitted and BVN copula with normal (upper panel graph) and beta (lower panel graph) margins. For normal margins, the axes are in  logit scale since  we also plot  the estimated contour plot of the random effects distribution as predictive region; this has been estimated for the logit pair of (Sensitivity, Specificity).}
\begin{center}
\begin{tabular}{|cc|}
\hline \multicolumn{2}{|c|}{Normal margins}\\
$^\dag$ BVN & Clayton \\\hline

\includegraphics[width=0.4\textwidth]{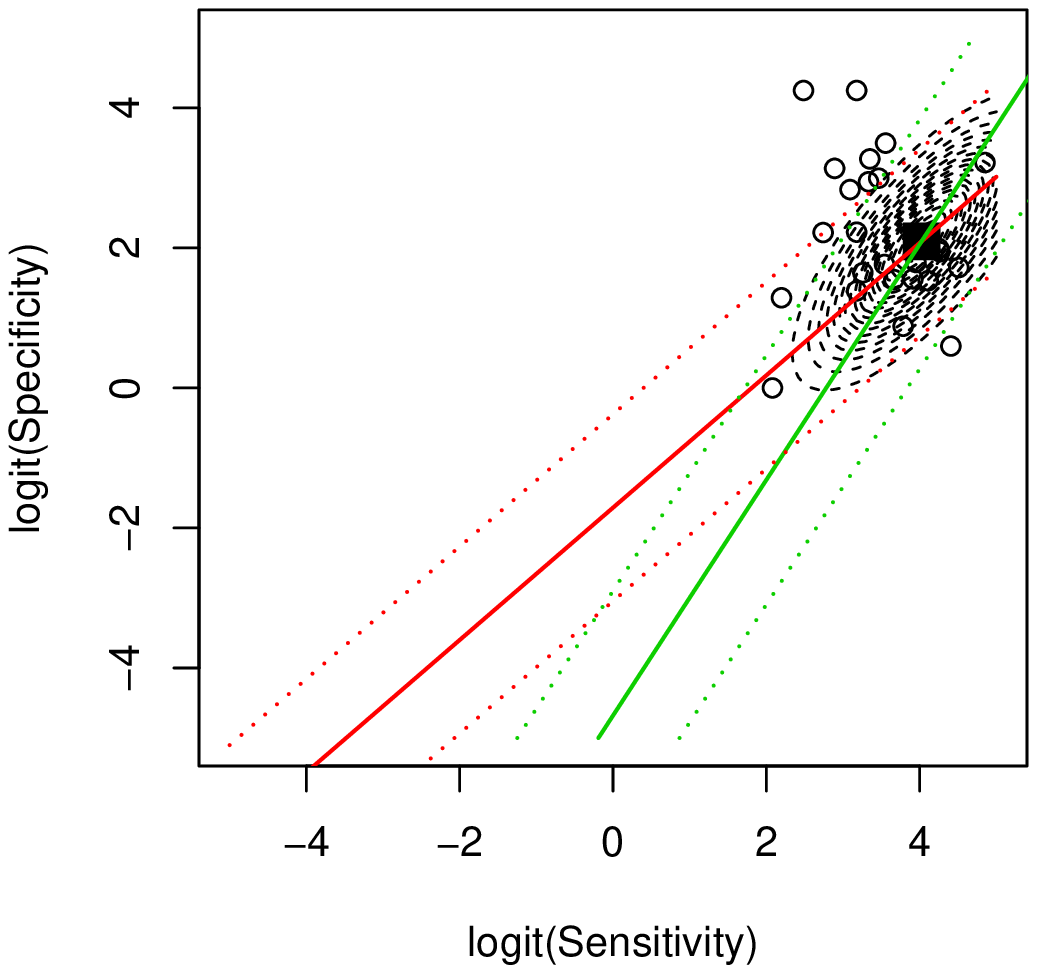}
&

\includegraphics[width=0.4\textwidth]{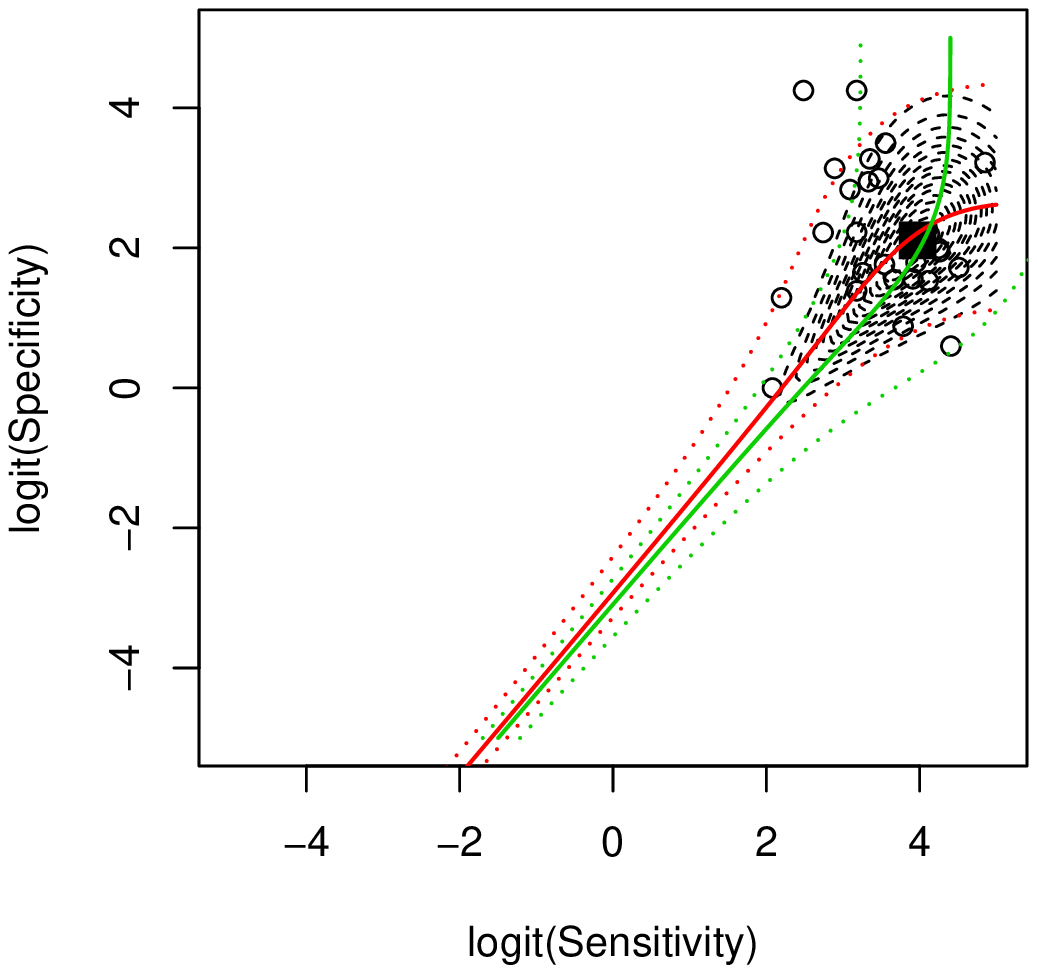}\\
\hline \multicolumn{2}{|c|}{Beta margins}\\
BVN & $^\S$ Clayton \\\hline
\includegraphics[width=0.4\textwidth]{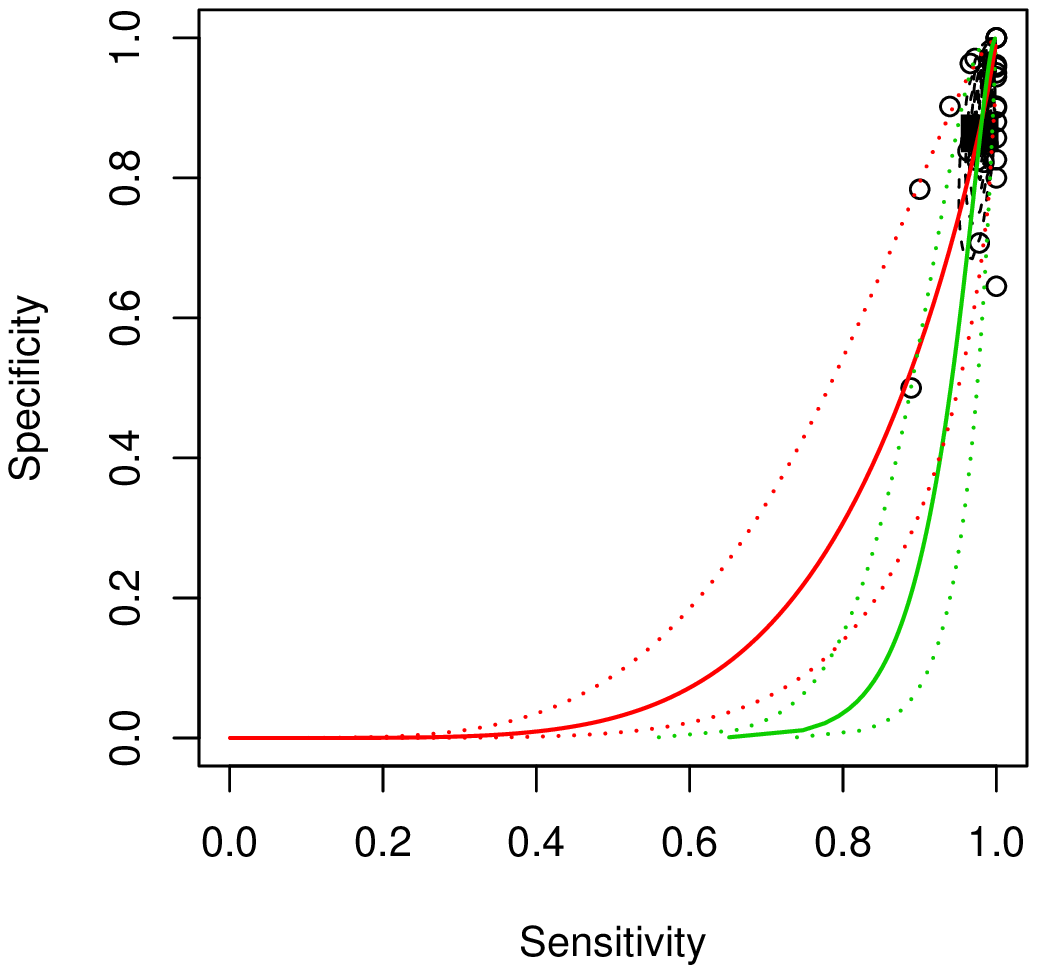}
&
\includegraphics[width=0.4\textwidth]{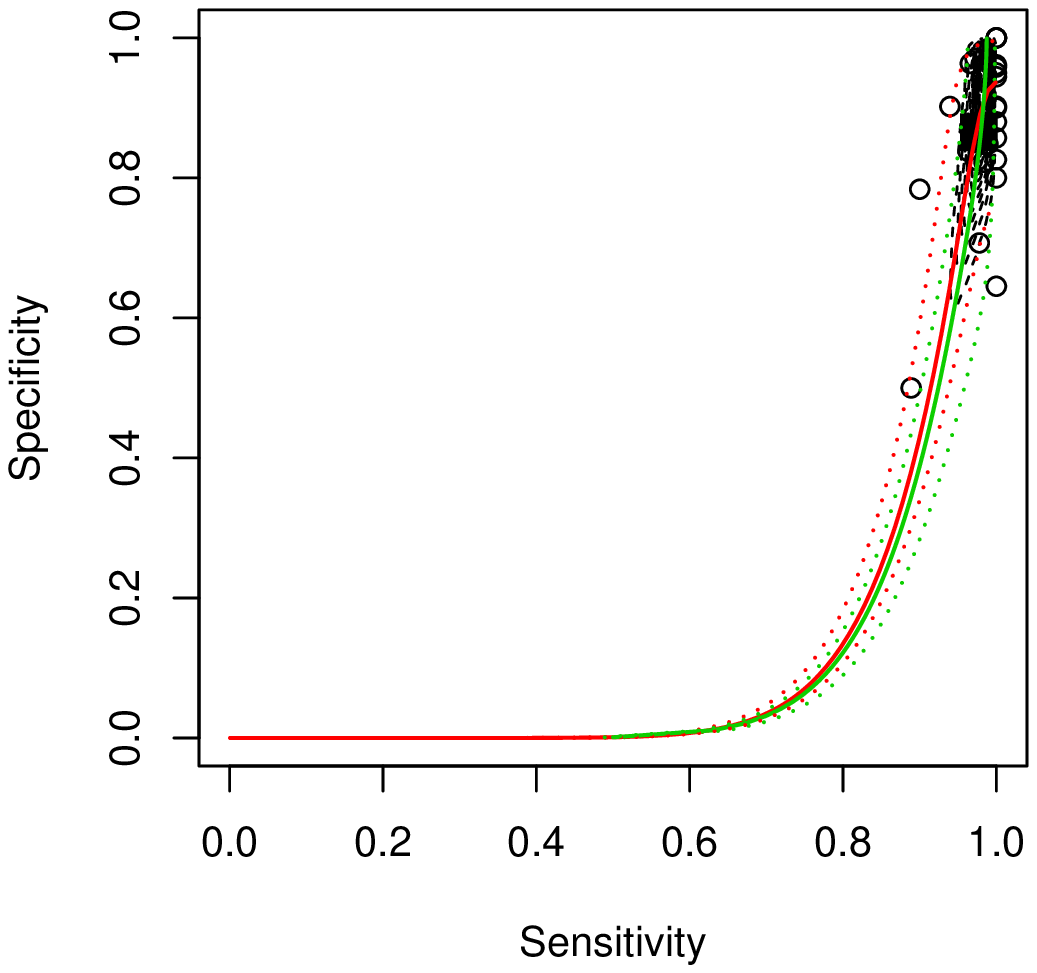}\\\hline
\end{tabular}

\end{center}
       \vspace{-1ex}
\begin{flushleft}
\begin{footnotesize}
$^\S$: Best fit;
$^\dag$: The resulting model is the same as the TGLMM;
$\blacksquare$: summary point; $\circ$: study estimate; 
Red and green lines represent the quantile  regression curves $x_1:=\widetilde{x}_1(x_2,q)$ and $x_2:=\widetilde{x}_2(x_1,q)$, respectively; for $q=0.5$ solid lines and for $q\in\{0.01,0.99\}$ dotted lines (confidence region). 
\end{footnotesize}  
\end{flushleft}
\end{figure}

Though typically the focus of meta-analysis has been to derive the  summary-effect estimates, there is increasing interest in drawing predictive inference. 
A  summary receiver operating characteristic (SROC) 
 curve has been  deduced for the  bivariate copula mixed model \cite{Nikoloulopoulos2015b} through a median regression  curve of $X_1$ on $X_2$. 
However, as there is no priori reason to regress $X_1$ on $X_2$ instead of the other way around,  Nikoloulopoulos \cite{Nikoloulopoulos2015b} has also provided  a median regression  curve of $X_2$ on $X_1$. 
In addition to using just median regression curves,  quantile regression curves with a focus on high ($q$ = 0.99) and low quantiles ($q$ = 0.01), which are strongly associated with the upper and lower tail dependence imposed from each parametric family of copulas, have also been proposed
\cite{Nikoloulopoulos2015b}. These can been seen as confidence regions of the median regression SROC curve.  Finally, a contour plot of the  the random effects distribution at the ML estimate has been proposed to preserve the nature of a bivariate response instead of a univariate response along with a covariate
\cite{Nikoloulopoulos2015b}. The contour plot can be seen as the predictive region of the estimated pair of sensitivity and specificity. The prediction region of the copula mixed model does not depend on the assumption of bivariate normality of the random effects as in the TGLMM and therefore  has a non-elliptical shape.

Figure \ref{SROCs} demonstrates the SROC curves and summary operating points (a pair of average sensitivity and specificity) with a confidence and a predictive region from the best fitted and BVN copula with normal (upper panel graph) and beta (lower panel graph) margins. From the graph it is apparent that  better prediction is achieved  when a Clayton copula with  beta margins  is assumed for the random effects distribution of the latent sensitivity and specificity.  

\section{\label{sec-discussion}Discussion}

We have exemplified the  vine copula mixed model for trivariate  meta-analysis of diagnostic test accuracy studies \cite{Nikoloulopoulos2015c} in the presence of non-evaluable subjects. 
It  includes the extended TGLMM \cite{ma-etal-2014} as a special case  and  it can be seen to provide an improvement over the latter  on the basis of the log-likelihood principle.  Hence,  superior statistical inference for the meta-analytic parameters of interest can be achieved when there is a belief in an MAR assumption.

This improvement relies on the
fact that the random effects distribution is expressed via vine copulas. 
The TVN distribution of the transformed latent proportions in the TGLMM  has restricted properties, i.e., a linear correlation structure and  normal margins.
Copulas  break the model building process into two separate steps, the  choice of arbitrary marginal distributions, and 
the choice of an arbitrary copula function  (dependence structure). Hence,  we can use beta instead of normal margins to model the latent proportions  in the original scale. The choice of the copula couldn't be other than the class of vine copulas.  Vine copulas 
allow for flexible tail dependence, different from assuming simple linear correlation structures, tail independence and normality \cite{joeetal10}, which makes them well suited for meta-analysis of diagnostic tests as the traditional assumption of multivariate normality is invalid.

It has been reported in the literature that in the TGLMM  estimation problems relating to the correlation parameters exist, such as non-convergence and a singular covariance matrix, particularly if the sample size is small \citep{chu-etal-2009}.
Nevertheless, we rather propose a numerically stable ML estimation technique based on Gauss-Legendre quadrature; the crucial step is to convert from independent to dependent quadrature points. 
The application example and simulations use a sufficient number of individual studies, i.e., $N=30$.  For meta-analyses with fewer studies one can simplify the model   using a truncated at level 1 vine copula. We refer the interested reader to   our previous study on trivariate vine copula mixed models  \cite{Nikoloulopoulos2015c} for simulations and various application examples that involve a small number of studies and call this notion of a truncated at level 1 vine copula.
The improvement over the reduction of the dependence parameters  is small (one dependence parameter less), but for estimation purposes this is extremely useful for a small number of studies.

In an era of evidence-based medicine, decision makers need high-quality procedures such as the SROC curves to support decisions about whether or not to use a diagnostic test in a specific clinical situation.
Different SROC curves essentially show  the
effect of different  model (random effect distribution) assumptions, since SROC is an inference that depends on the joint distribution.
For the vine copula mixed model, the model parameters (including dependence parameters), the choice of the pair copulas, and the choice of the margin affect the shape of the SROC curve \cite{Nikoloulopoulos2015b}, while  the SROC curve from the TGLMM  is severely restricted to the elliptical (linear) shape.

A recurrent theme underlying our methodology for analysis in the presence of missing data is the need to make assumptions that cannot be verified based on the observed data. Throughout this paper we adopted the  assumption of MAR. Nevertheless, it is natural to be concerned about robustness or sensitivity of inferences to departures from the MAR assumption. Future research will focus to handle the case  when the   non-evaluable subjects will be treated as  non-missing categories.

\section*{Software}
{\tt R} functions to derive  estimates  and simulate from  the vine copula mixed  model  
for trivariate meta-analysis of diagnostic studies in the presence of non-evaluable subjects are  part of the  {\tt  R} package {\tt  CopulaREMADA} \cite{Nikoloulopoulos-2018-CopulaREMADA}. The data and code  used in Section \ref{sec-app}  are given as code examples in the package.

\section*{Acknowledgements}
The simulations presented in this paper were carried out on the High Performance Computing Cluster supported by the Research and Specialist Computing Support service at the University of East Anglia.


\newpage

\begin{center}
\begin{Large}
{\bf  Supplementary Material for}
\end{Large}

\vspace{2cm}

{\LARGE
\bf An extended trivariate vine copula mixed model for  meta-analysis of diagnostic studies in the presence of non-evaluable outcomes}

\vspace{2cm}

{\large \bf Aristidis~K.~Nikoloulopoulos \footnote{\href{mailto:a.nikoloulopoulos@uea.ac.uk}{a.nikoloulopoulos@uea.ac.uk},  School of Computing Sciences, University of East Anglia, Norwich NR4 7TJ, U.K.} }

\end{center}

\thispagestyle{empty}
\setcounter{page}{0}
\setcounter{table}{0}

\setlength{\tabcolsep}{4pt}

\begin{table}[htbp]
  \centering
  \caption{\label{sim-norm-01.01}Biases,  root mean square errors (RMSE) and standard deviations (SD), along with the square root of the average theoretical variances ($\sqrt{\bar V}$), scaled by 100, for the MLEs  under different copula choices and margins from  $10^4$ small sample of sizes $N = 30$ simulations ($n_q=15$) from the trivariate vine copula mixed model with  Clayton  copulas  rotated by $90^\circ$  for both the $C_{12}(;\tau_{12})$ and $C_{13}(;\tau_{23|1})$ copulas, the Clayton copula for the $C_{13}(;\tau_{13})$ copula and normal margins. The missing probabilities for diseased and non-diseased subjects are the same, i.e.,  $v_4=v_5=0.1$.}
    \begin{tabular}{lllccccccccc}
    \toprule
          & margin & copula & $\pi_1$ & $\pi_2$ & $\pi_3$ & $\s_1$ & $\s_2$ & $\s_3$ & $\tau_{12}$ & $\tau_{13}$ & $\tau_{23|1}$ \\
    \midrule
  \rowcolor{gray}      Bias &$^\dag$  normal  & BVN   & 0.16  & -0.12 & 0.10  & -7.85 & -2.23 & -1.73 & -4.67 & -3.19 & 17.21 \\
          &       & Frank & 0.20  & -0.16 & -0.27 & -7.31 & -1.24 & -2.13 & -5.95 & -0.96 & 15.46 \\
   \rowcolor{gray}          &   $^\S$    & Cln$\{0^\circ,90^\circ\}$ & 0.30  & -0.05 & 0.90  & -6.62 & 2.00  & -3.94 & -3.64 & -6.93 & 20.98 \\
          &       & Cln$\{0^\circ,270^\circ\}$ & 1.25  & -0.24 & 0.04  & -9.21 & -1.58 & 1.69  & 5.31  & -16.72 & 17.55 \\
   \rowcolor{gray}          & beta  & BVN   & -2.09 & -3.59 & 4.05  & -     & -     & -     & -4.02 & -4.53 & 19.03 \\
          &       & Frank & -1.80 & -3.56 & 3.55  & -     & -     & -     & -5.86 & -1.62 & 17.51 \\
   \rowcolor{gray}          &      & Cln$\{0^\circ,90^\circ\}$ & -1.82 & -3.76 & 4.52  & -     & -     & -     & -2.67 & -10.09 & 19.11 \\
          &       & Cln$\{0^\circ,270^\circ\}$ & -1.15 & -3.75 & 4.22  & -     & -     & -     & 5.06  & -16.89 & 20.57 \\\hline
   \rowcolor{gray}     SD &$^\dag$ normal & BVN   & 4.34  & 1.85  & 3.54  & 17.82 & 16.43 & 14.91 & 15.14 & 13.23 & 23.63 \\
          &       & Frank & 4.61  & 1.93  & 3.59  & 18.57 & 17.18 & 14.98 & 16.33 & 14.19 & 24.25 \\
   \rowcolor{gray}          &  $^\S$      & Cln$\{0^\circ,90^\circ\}$ & 4.51  & 1.91  & 3.79  & 19.69 & 18.66 & 15.77 & 13.85 & 17.67 & 24.59 \\
          &       & Cln$\{0^\circ,270^\circ\}$ & 4.29  & 2.00  & 3.72  & 18.96 & 18.44 & 17.75 & 26.62 & 17.98 & 22.34 \\
    \rowcolor{gray}         & beta  & BVN   & 3.93  & 2.27  & 3.36  & 3.97  & 3.00  & 3.38  & 14.80 & 13.22 & 23.57 \\
          &       & Frank & 4.12  & 2.33  & 3.36  & 4.17  & 3.07  & 3.36  & 16.02 & 14.04 & 24.82 \\
    \rowcolor{gray}         &     & Cln$\{0^\circ,90^\circ\}$ & 4.09  & 2.45  & 3.63  & 4.40  & 3.72  & 3.55  & 13.94 & 17.42 & 25.30 \\
          &       & Cln$\{0^\circ,270^\circ\}$ & 3.90  & 2.41  & 3.61  & 4.23  & 3.18  & 4.24  & 26.48 & 18.94 & 22.44 \\\hline
    \rowcolor{gray}    $\sqrt{\bar V}$ &$^\dag$ normal & BVN   & 3.96  & 1.75  & 3.36  & 16.77 & 15.49 & 13.55 & 12.37 & 11.02 & 18.24 \\
          &       & Frank & 3.88  & 1.72  & 3.18  & 17.00 & 15.72 & 13.16 & 12.01 & 10.82 & 17.07 \\
    \rowcolor{gray}         & $^\S$      & Cln$\{0^\circ,90^\circ\}$ & 3.74  & 1.67  & 2.82  & 16.10 & 14.39 & 10.77 & 10.93 & 8.33  & 13.22 \\
          &       & Cln$\{0^\circ,270^\circ\}$ & 3.39  & 1.62  & 2.98  & 15.77 & 14.11 & 13.03 & 11.19 & 7.60  & 10.05 \\
    \rowcolor{gray}         & beta  & BVN   & 3.54  & 1.98  & 3.04  & 3.77  & 2.49  & 3.02  & 12.46 & 11.15 & 17.99 \\
          &       & Frank & 3.52  & 1.92  & 2.89  & 3.88  & 2.48  & 2.88  & 12.42 & 10.92 & 17.50 \\
  \rowcolor{gray}           &       & Cln$\{0^\circ,90^\circ\}$ & 3.40  & 1.81  & 2.47  & 3.59  & 2.41  & 2.30  & 11.21 & 8.74  & 13.76 \\
          &       & Cln$\{0^\circ,270^\circ\}$ & 3.13  & 1.80  & 2.70  & 3.51  & 2.24  & 2.84  & 11.68 & 8.08  & 10.21 \\\hline
  \rowcolor{gray}      RMSE &$^\dag$ normal & BVN   & 4.34  & 1.85  & 3.54  & 19.47 & 16.58 & 15.01 & 15.84 & 13.60 & 29.23 \\
          &       & Frank & 4.62  & 1.93  & 3.60  & 19.96 & 17.23 & 15.13 & 17.38 & 14.22 & 28.76 \\
  \rowcolor{gray}           & $^\S$      & Cln$\{0^\circ,90^\circ\}$ & 4.52  & 1.91  & 3.90  & 20.78 & 18.77 & 16.26 & 14.32 & 18.98 & 32.32 \\
          &       & Cln$\{0^\circ,270^\circ\}$ & 4.47  & 2.01  & 3.72  & 21.08 & 18.50 & 17.83 & 27.15 & 24.55 & 28.41 \\
  \rowcolor{gray}           & beta  & BVN   & 4.45  & 4.24  & 5.26  & -     & -     & -     & 15.34 & 13.97 & 30.29 \\
          &       & Frank & 4.49  & 4.26  & 4.89  & -     & -     & -     & 17.06 & 14.14 & 30.37 \\
  \rowcolor{gray}           &      & Cln$\{0^\circ,90^\circ\}$ & 4.48  & 4.49  & 5.80  & -     & -     & -     & 14.19 & 20.13 & 31.71 \\
          &       & Cln$\{0^\circ,270^\circ\}$ & 4.06  & 4.46  & 5.55  & -     & -     & -     & 26.96 & 25.37 & 30.44 \\
    \bottomrule
    \end{tabular}%
      \vspace{-1ex}
\begin{flushleft}
\begin{footnotesize}
$^\S$: True model;
$^\dag$: The resulting model is the same as the TGLMM; Cln\{$\omega_1^\circ,\omega_2^\circ$\}: The $C_{13}(\cdot;\tau_{13})$ and \{$C_{12}(\cdot;\tau_{12}), C_{23|1}(\cdot;\tau_{23|1})$\} pair copulas are Clayton rotated by $\omega_1$ and $\omega_2$ degrees, respectively.
\end{footnotesize}  
\end{flushleft}
  \label{tab:addlabel}%
\end{table}

\begin{table}[htbp]
  \centering
 \caption{\label{sim-norm-02.01}Biases,  root mean square errors (RMSE) and standard deviations (SD), along with the square root of the average theoretical variances ($\sqrt{\bar V}$), scaled by 100, for the MLEs  under different copula choices and margins from  $10^4$ small sample of sizes $N = 30$ simulations ($n_q=15$) from the trivariate vine copula mixed model with  Clayton  copulas  rotated by $90^\circ$  for both the $C_{12}(;\tau_{12})$ and $C_{13}(;\tau_{23|1})$ copulas, the Clayton copula for the $C_{13}(;\tau_{13})$ copula and normal margins.  The missing probability of diseased group is larger than non-diseased group, i.e.,  $v_4=0.2>v_5=0.1$.}
    \begin{tabular}{lllccccccccc}
    \toprule
          & margin & copula & $\pi_1$ & $\pi_2$ & $\pi_3$ & $\s_1$ & $\s_2$ & $\s_3$ & $\tau_{12}$ & $\tau_{13}$ & $\tau_{23|1}$ \\
    \midrule
  \rowcolor{gray}      Bias &$^\dag$ normal & BVN   & 0.23  & -0.12 & 0.10  & -8.23 & -2.37 & -1.92 & -4.97 & -3.36 & 17.51 \\
          &       & Frank & 0.27  & -0.16 & -0.25 & -7.81 & -1.46 & -2.34 & -6.30 & -1.07 & 15.89 \\
  \rowcolor{gray}           & $^\S$      & Cln$\{0^\circ,90^\circ\}$ & 0.38  & -0.05 & 0.88  & -7.15 & 1.97  & -3.90 & -3.75 & -6.58 & 20.99 \\
          &       & Cln$\{0^\circ,270^\circ\}$ & 1.33  & -0.25 & 0.08  & -9.77 & -1.73 & 1.39  & 4.01  & -16.30 & 17.06 \\
          & beta  & BVN   & -1.95 & -3.60 & 4.07  & -     & -     & -     & -4.30 & -4.84 & 19.04 \\
          &       & Frank & -1.66 & -3.58 & 3.60  & -     & -     & -     & -6.18 & -1.87 & 18.03 \\
     \rowcolor{gray}        &    & Cln$\{0^\circ,90^\circ\}$ & -1.68 & -3.75 & 4.50  & -     & -     & -     & -2.71 & -9.87 & 18.86 \\
          &       & Cln$\{0^\circ,270^\circ\}$ & -1.01 & -3.78 & 4.24  & -     & -     & -     & 3.90  & -16.65 & 19.97 \\\hline
     SD &$^\dag$ normal & BVN   & 4.40  & 1.85  & 3.56  & 18.37 & 16.30 & 14.91 & 15.49 & 13.48 & 23.90 \\
          &       & Frank & 4.64  & 1.92  & 3.59  & 19.06 & 16.96 & 14.99 & 16.81 & 14.49 & 25.15 \\
     \rowcolor{gray}        & $^\S$      & Cln$\{0^\circ,90^\circ\}$ & 4.61  & 1.92  & 3.81  & 20.28 & 18.54 & 15.72 & 14.08 & 18.15 & 24.63 \\
          &       & Cln$\{0^\circ,270^\circ\}$ & 4.30  & 2.00  & 3.74  & 19.55 & 18.17 & 17.63 & 27.34 & 18.38 & 22.75 \\
    \rowcolor{gray}         & beta  & BVN   & 3.97  & 2.27  & 3.36  & 4.07  & 2.99  & 3.38  & 15.25 & 13.49 & 23.76 \\
          &       & Frank & 4.13  & 2.33  & 3.37  & 4.25  & 3.04  & 3.36  & 16.69 & 14.45 & 25.94 \\
     \rowcolor{gray}        &     & Cln$\{0^\circ,90^\circ\}$ & 4.16  & 2.45  & 3.64  & 4.44  & 3.70  & 3.54  & 14.29 & 17.92 & 25.75 \\
          &       & Cln$\{0^\circ,270^\circ\}$ & 3.92  & 2.43  & 3.61  & 4.34  & 3.17  & 4.18  & 27.32 & 19.34 & 22.54 \\\hline
    \rowcolor{gray}    $\sqrt{\bar V}$ &$^\dag$ normal & BVN   & 4.03  & 1.75  & 3.37  & 17.28 & 15.51 & 13.54 & 12.79 & 11.31 & 18.65 \\
          &       & Frank & 3.94  & 1.72  & 3.19  & 17.47 & 15.70 & 13.15 & 12.29 & 11.09 & 17.52 \\
    \rowcolor{gray}         & $^\S$      & Cln$\{0^\circ,90^\circ\}$ & 3.80  & 1.67  & 2.83  & 16.57 & 14.39 & 10.81 & 11.09 & 8.44  & 13.48 \\
          &       & Cln$\{0^\circ,270^\circ\}$ & 3.46  & 1.63  & 3.00  & 16.30 & 14.13 & 12.99 & 11.20 & 7.76  & 10.23 \\
     \rowcolor{gray}        & beta  & BVN   & 3.58  & 1.98  & 3.05  & 3.88  & 2.49  & 3.01  & 12.92 & 11.44 & 18.54 \\
          &       & Frank & 3.56  & 1.92  & 2.89  & 3.99  & 2.47  & 2.87  & 12.74 & 11.23 & 17.82 \\
      \rowcolor{gray}       &       & Cln$\{0^\circ,90^\circ\}$ & 3.45  & 1.81  & 2.47  & 3.66  & 2.39  & 2.29  & 11.44 & 8.84  & 13.93 \\
          &       & Cln$\{0^\circ,270^\circ\}$ & 3.18  & 1.80  & 2.71  & 3.62  & 2.23  & 2.84  & 11.77 & 8.23  & 10.27 \\\hline
     \rowcolor{gray}   RMSE &$^\dag$ normal & BVN   & 4.40  & 1.86  & 3.56  & 20.13 & 16.47 & 15.04 & 16.27 & 13.90 & 29.63 \\
          &       & Frank & 4.65  & 1.93  & 3.60  & 20.60 & 17.02 & 15.17 & 17.95 & 14.53 & 29.75 \\
     \rowcolor{gray}        & $^\S$      & Cln$\{0^\circ,90^\circ\}$ & 4.62  & 1.92  & 3.92  & 21.51 & 18.65 & 16.20 & 14.57 & 19.31 & 32.36 \\
          &       & Cln$\{0^\circ,270^\circ\}$ & 4.50  & 2.01  & 3.74  & 21.86 & 18.25 & 17.68 & 27.63 & 24.57 & 28.44 \\
    \rowcolor{gray}         & beta  & BVN   & 4.43  & 4.25  & 5.28  & -     & -     & -     & 15.84 & 14.33 & 30.45 \\
          &       & Frank & 4.45  & 4.27  & 4.93  & -     & -     & -     & 17.80 & 14.58 & 31.59 \\
     \rowcolor{gray}        &    & Cln$\{0^\circ,90^\circ\}$ & 4.49  & 4.48  & 5.79  & -     & -     & -     & 14.54 & 20.46 & 31.92 \\
          &       & Cln$\{0^\circ,270^\circ\}$ & 4.05  & 4.49  & 5.57  & -     & -     & -     & 27.59 & 25.52 & 30.11 \\
    \bottomrule
    \end{tabular}%
     \vspace{-1ex}
\begin{flushleft}
\begin{footnotesize}
$^\S$: True model;
$^\dag$: The resulting model is the same as the TGLMM; Cln\{$\omega_1^\circ,\omega_2^\circ$\}: The $C_{13}(\cdot;\tau_{13})$ and \{$C_{12}(\cdot;\tau_{12}), C_{23|1}(\cdot;\tau_{23|1})$\} pair copulas are Clayton rotated by $\omega_1$ and $\omega_2$ degrees, respectively.
\end{footnotesize}  
\end{flushleft}
  \label{tab:addlabel}%
\end{table}

\begin{table}[htbp]
  \centering
  \caption{\label{sim-beta-01.01}Biases,  root mean square errors (RMSE) and standard deviations (SD), along with the square root of the average theoretical variances ($\sqrt{\bar V}$), scaled by 100, for the MLEs  under different copula choices and margins from  $10^4$ small sample of sizes $N = 30$ simulations ($n_q=15$) from the trivariate vine copula mixed model with  Clayton  copulas  rotated by $90^\circ$  for both the $C_{12}(;\tau_{12})$ and $C_{13}(;\tau_{23|1})$ copulas, the Clayton copula for the $C_{13}(;\tau_{13})$ copula and beta margins. The missing probabilities for diseased and non-diseased subjects are the same, i.e.,  $v_4=v_5=0.1$.}
    \begin{tabular}{lllccccccccc}
    \toprule
          & margin & copula & $\pi_1$ & $\pi_2$ & $\pi_3$ & $\g_1$ & $\g_2$ & $\g_3$ & $\tau_{12}$ & $\tau_{13}$ & $\tau_{23|1}$ \\
    \midrule
  \rowcolor{gray}      Bias &$^\dag$ normal & BVN   & 2.30  & 3.59  & -2.78 & -     & -     & -     & -5.71 & -2.61 & 14.54 \\
          &       & Frank & 2.17  & 3.51  & -2.89 & -     & -     & -     & -6.96 & -0.70 & 12.33 \\
    \rowcolor{gray}         &       & Cln$\{0^\circ,90^\circ\}$ & 2.22  & 3.67  & -2.52 & -     & -     & -     & -4.04 & -2.62 & 15.72 \\
          &       & Cln$\{0^\circ,270^\circ\}$ & 2.98  & 3.51  & -2.79 & -     & -     & -     & 1.31  & -12.10 & 10.89 \\
   \rowcolor{gray}          & beta  & BVN   & 0.65  & -0.02 & 0.10  & -1.50 & -0.42 & -0.57 & -6.48 & -3.65 & 18.12 \\
          &       & Frank & 0.73  & -0.12 & -0.12 & -1.52 & -0.23 & -0.69 & -7.85 & -0.99 & 17.17 \\
   \rowcolor{gray}          & $^\S$      & Cln$\{0^\circ,90^\circ\}$ & 0.72  & -0.02 & 0.21  & -1.54 & -0.09 & -1.04 & -3.92 & -5.90 & 15.13 \\
          &       & Cln$\{0^\circ,270^\circ\}$ & 1.35  & -0.24 & 0.21  & -1.95 & -0.26 & -0.13 & -0.33 & -12.28 & 15.11 \\\hline
   \rowcolor{gray}     SD &$^\dag$ normal & BVN   & 3.50  & 1.59  & 2.77  & 16.26 & 20.91 & 13.08 & 17.04 & 15.65 & 30.27 \\
          &       & Frank & 3.67  & 1.65  & 2.81  & 16.71 & 22.07 & 13.12 & 18.44 & 16.95 & 31.64 \\
   \rowcolor{gray}          &       & Cln$\{0^\circ,90^\circ\}$ & 3.68  & 1.64  & 2.91  & 17.45 & 21.86 & 13.91 & 14.80 & 20.88 & 28.19 \\
          &       & Cln$\{0^\circ,270^\circ\}$ & 3.44  & 1.71  & 2.86  & 17.13 & 23.36 & 15.00 & 28.26 & 20.83 & 29.38 \\
    \rowcolor{gray}         & beta  & BVN   & 3.28  & 1.87  & 2.61  & 3.16  & 2.91  & 2.42  & 17.13 & 15.64 & 31.02 \\
          &       & Frank & 3.40  & 1.97  & 2.62  & 3.25  & 3.08  & 2.39  & 18.87 & 16.91 & 32.42 \\
      \rowcolor{gray}       & $^\S$      & Cln$\{0^\circ,90^\circ\}$ & 3.41  & 1.95  & 2.74  & 3.32  & 3.19  & 2.46  & 14.97 & 20.90 & 28.71 \\
          &       & Cln$\{0^\circ,270^\circ\}$ & 3.21  & 2.02  & 2.70  & 3.27  & 3.20  & 2.85  & 28.91 & 22.01 & 29.84 \\\hline
    \rowcolor{gray}    $\sqrt{\bar V}$ &$^\dag$ normal & BVN   & 3.25  & 1.42  & 2.63  & 14.95 & 20.02 & 11.74 & 14.03 & 13.32 & 23.34 \\
          &       & Frank & 3.20  & 1.42  & 2.53  & 14.90 & 20.51 & 11.56 & 12.72 & 13.04 & 20.91 \\
    \rowcolor{gray}         &      & Cln$\{0^\circ,90^\circ\}$ & 3.13  & 1.34  & 2.34  & 14.48 & 18.15 & 10.00 & 11.35 & 10.33 & 15.90 \\
          &       & Cln$\{0^\circ,270^\circ\}$ & 2.87  & 1.35  & 2.48  & 14.04 & 18.83 & 11.58 & 10.43 & 10.07 & 13.53 \\
    \rowcolor{gray}         & beta  & BVN   & 3.05  & 1.78  & 2.48  & 3.01  & 2.77  & 2.30  & 14.45 & 13.43 & 24.36 \\
          &       & Frank & 3.03  & 1.77  & 2.39  & 3.01  & 2.86  & 2.23  & 13.67 & 13.25 & 22.08 \\
    \rowcolor{gray}         & $^\S$      & Cln$\{0^\circ,90^\circ\}$ & 2.98  & 1.60  & 2.15  & 2.87  & 2.52  & 1.87  & 12.01 & 10.94 & 16.58 \\
          &       & Cln$\{0^\circ,270^\circ\}$ & 2.77  & 1.68  & 2.39  & 2.78  & 2.63  & 2.29  & 11.70 & 10.87 & 14.47 \\\hline
     \rowcolor{gray}   RMSE &$^\dag$ normal & BVN   & 4.19  & 3.93  & 3.92  & -     & -     & -     & 17.97 & 15.87 & 33.58 \\
          &       & Frank & 4.27  & 3.88  & 4.03  & -     & -     & -     & 19.71 & 16.96 & 33.96 \\
     \rowcolor{gray}        &      & Cln$\{0^\circ,90^\circ\}$ & 4.30  & 4.02  & 3.85  & -     & -     & -     & 15.35 & 21.05 & 32.28 \\
          &       & Cln$\{0^\circ,270^\circ\}$ & 4.55  & 3.90  & 3.99  & -     & -     & -     & 28.29 & 24.09 & 31.33 \\
     \rowcolor{gray}        & beta  & BVN   & 3.34  & 1.87  & 2.61  & 3.50  & 2.94  & 2.48  & 18.32 & 16.07 & 35.93 \\
          &       & Frank & 3.48  & 1.97  & 2.62  & 3.59  & 3.09  & 2.49  & 20.44 & 16.94 & 36.69 \\
     \rowcolor{gray}        & $^\S$      & Cln$\{0^\circ,90^\circ\}$ & 3.48  & 1.95  & 2.75  & 3.66  & 3.19  & 2.67  & 15.47 & 21.72 & 32.45 \\
          &       & Cln$\{0^\circ,270^\circ\}$ & 3.48  & 2.03  & 2.71  & 3.81  & 3.21  & 2.85  & 28.91 & 25.21 & 33.45 \\
    \bottomrule
    \end{tabular}%
     \vspace{-1ex}
\begin{flushleft}
\begin{footnotesize}
$^\S$: True model;
$^\dag$: The resulting model is the same as the TGLMM; Cln\{$\omega_1^\circ,\omega_2^\circ$\}: The $C_{13}(\cdot;\tau_{13})$ and \{$C_{12}(\cdot;\tau_{12}), C_{23|1}(\cdot;\tau_{23|1})$\} pair copulas are Clayton rotated by $\omega_1$ and $\omega_2$ degrees, respectively.
\end{footnotesize}  
\end{flushleft}
  \label{tab:addlabel}%
\end{table}

\begin{table}[htbp]
  \centering
 \caption{\label{sim-beta-02.01}Biases,  root mean square errors (RMSE) and standard deviations (SD), along with the square root of the average theoretical variances ($\sqrt{\bar V}$), scaled by 100, for the MLEs  under different copula choices and margins from  $10^4$ small sample of sizes $N = 30$ simulations ($n_q=15$) from the trivariate vine copula mixed model with  Clayton  copulas  rotated by $90^\circ$  for both the $C_{12}(;\tau_{12})$ and $C_{13}(;\tau_{23|1})$ copulas, the Clayton copula for the $C_{13}(;\tau_{13})$ copula and beta margins.  The missing probability of diseased group is larger than non-diseased group, i.e.,  $v_4=0.2>v_5=0.1$.}
    \begin{tabular}{lllccccccccc}
    \toprule
          & margin & copula & $\pi_1$ & $\pi_2$ & $\pi_3$ & $\g_1$ & $\g_2$ & $\g_3$ & $\tau_{12}$ & $\tau_{13}$ & $\tau_{23|1}$ \\
    \midrule
  \rowcolor{gray}      Bias &$^\dag$ normal & BVN   & 2.31  & 3.61  & -2.79 & -     & -     & -     & -6.23 & -2.45 & 14.65 \\
          &       & Frank & 2.20  & 3.54  & -2.90 & -     & -     & -     & -7.42 & -0.79 & 12.08 \\
    \rowcolor{gray}         &       & Cln$\{0^\circ,90^\circ\}$ & 2.22  & 3.70  & -2.55 & -     & -     & -     & -4.31 & -2.08 & 15.57 \\
          &       & Cln$\{0^\circ,270^\circ\}$ & 2.99  & 3.54  & -2.79 & -     & -     & -     & 0.35  & -11.53 & 10.19 \\
     \rowcolor{gray}        & beta  & BVN   & 0.72  & -0.01 & 0.10  & -1.54 & -0.40 & -0.62 & -6.84 & -3.68 & 17.82 \\
          &       & Frank & 0.81  & -0.11 & -0.11 & -1.54 & -0.22 & -0.73 & -8.34 & -1.09 & 16.99 \\
     \rowcolor{gray}        & $^\S$      & Cln$\{0^\circ,90^\circ\}$ & 0.79  & 0.00  & 0.20  & -1.60 & -0.08 & -1.05 & -3.90 & -5.64 & 15.04 \\
          &       & Cln$\{0^\circ,270^\circ\}$ & 1.41  & -0.23 & 0.23  & -1.99 & -0.23 & -0.23 & -0.77 & -11.75 & 14.12 \\\hline
   \rowcolor{gray}     SD &$^\dag$ normal & BVN   & 3.55  & 1.58  & 2.78  & 16.73 & 20.95 & 13.20 & 17.76 & 16.29 & 31.53 \\
          &       & Frank & 3.70  & 1.65  & 2.80  & 17.14 & 22.15 & 13.31 & 19.18 & 17.27 & 31.96 \\
  \rowcolor{gray}           &      & Cln$\{0^\circ,90^\circ\}$ & 3.74  & 1.64  & 2.92  & 17.87 & 22.03 & 14.09 & 15.23 & 21.30 & 28.90 \\
          &       & Cln$\{0^\circ,270^\circ\}$ & 3.48  & 1.70  & 2.86  & 17.51 & 23.51 & 14.93 & 28.85 & 21.60 & 29.88 \\
  \rowcolor{gray}           & beta  & BVN   & 3.31  & 1.86  & 2.61  & 3.24  & 2.91  & 2.43  & 18.09 & 16.45 & 32.47 \\
          &       & Frank & 3.43  & 1.97  & 2.62  & 3.33  & 3.08  & 2.40  & 19.93 & 17.41 & 33.82 \\
   \rowcolor{gray}          & $^\S$      & Cln$\{0^\circ,90^\circ\}$ & 3.45  & 1.95  & 2.75  & 3.38  & 3.20  & 2.47  & 15.43 & 21.32 & 29.24 \\
          &       & Cln$\{0^\circ,270^\circ\}$ & 3.24  & 2.02  & 2.72  & 3.35  & 3.25  & 2.80  & 29.75 & 22.91 & 29.87 \\\hline
   \rowcolor{gray}     $\sqrt{\bar V}$ &$^\dag$ normal & BVN   & 3.31  & 1.41  & 2.62  & 15.49 & 20.03 & 11.73 & 14.52 & 13.76 & 24.73 \\
          &       & Frank & 3.26  & 1.41  & 2.53  & 15.41 & 20.53 & 11.56 & 12.93 & 13.30 & 21.08 \\
   \rowcolor{gray}          &       & Cln$\{0^\circ,90^\circ\}$ & 3.21  & 1.34  & 2.35  & 15.03 & 18.28 & 10.06 & 11.59 & 10.53 & 16.16 \\
          &       & Cln$\{0^\circ,270^\circ\}$ & 2.91  & 1.34  & 2.47  & 14.57 & 18.87 & 11.53 & 10.40 & 10.23 & 13.70 \\
   \rowcolor{gray}          & beta  & BVN   & 3.11  & 1.78  & 2.48  & 3.11  & 2.77  & 2.29  & 15.05 & 13.90 & 25.80 \\
          &       & Frank & 3.09  & 1.77  & 2.39  & 3.11  & 2.86  & 2.22  & 14.04 & 13.54 & 22.91 \\
   \rowcolor{gray}          & $^\S$      & Cln$\{0^\circ,90^\circ\}$ & 3.04  & 1.61  & 2.16  & 2.96  & 2.54  & 1.88  & 12.29 & 11.23 & 17.11 \\
          &       & Cln$\{0^\circ,270^\circ\}$ & 2.83  & 1.68  & 2.39  & 2.88  & 2.66  & 2.27  & 11.91 & 11.19 & 14.98 \\\hline
  \rowcolor{gray}      RMSE &$^\dag$ normal & BVN   & 4.23  & 3.94  & 3.93  & -     & -     & -     & 18.83 & 16.48 & 34.76 \\
          &       & Frank & 4.30  & 3.90  & 4.03  & -     & -     & -     & 20.56 & 17.29 & 34.17 \\
  \rowcolor{gray}           &     & Cln$\{0^\circ,90^\circ\}$ & 4.35  & 4.05  & 3.88  & -     & -     & -     & 15.83 & 21.40 & 32.83 \\
          &       & Cln$\{0^\circ,270^\circ\}$ & 4.58  & 3.92  & 3.99  & -     & -     & -     & 28.85 & 24.48 & 31.57 \\
   \rowcolor{gray}          & beta  & BVN   & 3.38  & 1.86  & 2.62  & 3.59  & 2.94  & 2.50  & 19.34 & 16.85 & 37.04 \\
          &       & Frank & 3.52  & 1.97  & 2.63  & 3.67  & 3.09  & 2.51  & 21.60 & 17.45 & 37.85 \\
   \rowcolor{gray}          & $^\S$      & Cln$\{0^\circ,90^\circ\}$ & 3.54  & 1.95  & 2.76  & 3.75  & 3.20  & 2.68  & 15.91 & 22.05 & 32.88 \\
          &       & Cln$\{0^\circ,270^\circ\}$ & 3.53  & 2.03  & 2.73  & 3.90  & 3.26  & 2.81  & 29.76 & 25.74 & 33.04 \\
    \bottomrule
    \end{tabular}%
     \vspace{-1ex}
\begin{flushleft}
\begin{footnotesize}
$^\S$: True model;
$^\dag$: The resulting model is the same as the TGLMM; Cln\{$\omega_1^\circ,\omega_2^\circ$\}: The $C_{13}(\cdot;\tau_{13})$ and \{$C_{12}(\cdot;\tau_{12}), C_{23|1}(\cdot;\tau_{23|1})$\} pair copulas are Clayton rotated by $\omega_1$ and $\omega_2$ degrees, respectively.
\end{footnotesize}  
\end{flushleft}
  \label{tab:addlabel}%
\end{table}

\end{document}